\documentclass[11pt]{article}
\pdfoutput=1

\usepackage{jheppub}
\usepackage[utf8]{inputenc}
\usepackage{booktabs}
\usepackage{graphicx}
\usepackage{subcaption}
\usepackage{mathtools}
\usepackage{cancel}
\usepackage{datetime}
\usepackage[dvipsnames]{xcolor}
\usepackage{slashed}
\usepackage{tcolorbox}
\usepackage{multirow}
\numberwithin{equation}{section}

\def\ash{%
  {
    a%
    \kern-.045em%
    \lower.3ex\hbox{sh}%
}%
}

\newcommand{\nn}{\nonumber}

\makeatletter
\newenvironment{tsubarray}[1]{%
  \vcenter\bgroup
  \Let@ \restore@math@cr \default@tag
  \baselineskip\fontdimen10 \scriptfont\tw@
  \advance\baselineskip\fontdimen12 \scriptfont\tw@
  \lineskip\thr@@\fontdimen8 \scriptfont\thr@@
  \lineskiplimit\lineskip
  \check@mathfonts
  \ialign\bgroup\ifx c#1\hfil\fi
    \normalfont\fontsize\sf@size\z@\selectfont\ignorespaces##\unskip\hfil\crcr
}{%
  \crcr\egroup\egroup
}
\makeatother

\makeatletter
\newcommand{\subalign}[1]{%
  \vcenter{%
    \Let@ \restore@math@cr \default@tag
    \baselineskip\fontdimen10 \scriptfont\tw@
    \advance\baselineskip\fontdimen12 \scriptfont\tw@
    \lineskip\thr@@\fontdimen8 \scriptfont\thr@@
    \lineskiplimit\lineskip
    \ialign{\hfil$\m@th\scriptstyle##$&$\m@th\scriptstyle{}##$\hfil\crcr
      #1\crcr
    }%
  }%
}
\makeatother

\newcommand{\be}{\begin{equation}}
\newcommand{\ee}{\end{equation}}
\newcommand\beq{\begin{equation}}
\newcommand\eeq{\end{equation}}
\newcommand\beal{\begin{aligned}}
\newcommand\eeal{\end{aligned}}
\newcommand\bea{\begin{eqnarray}}
\newcommand\eea{\end{eqnarray}}
\newcommand{\mbf}[1]{\mathbf{#1}}

\usepackage{xcolor}

\makeatletter
\newcommand{\Biggg}{\bBigg@{3.5}}
\makeatother

\begin{document}
\title{Full-shape BOSS constraints on dark matter interacting with dark radiation and lifting the $S_8$ tension}
\author{Henrique Rubira$^1$,}
\author{Asmaa Mazoun$^{1,2}$}
\author{and Mathias Garny$^2$}
\affiliation{\small $^1$ Physik Department T31, Technische Universit\"at M\"unchen \\
James-Franck-Stra\ss e 1, D-85748 Garching, Germany}
\affiliation{\small $^2$ University Observatory, Faculty of Physics, Ludwig-Maximilians-Universität\\ Scheinerstra\ss e 1, D-81679 München, Germany}
\emailAdd{henrique.rubira@tum.de, asmaa.mazoun@tum.de, mathias.garny@tum.de}

\preprint{TUM-HEP-1414/22}

\abstract{In this work we derive constraints on interacting dark matter-dark radiation models from a full-shape analysis of BOSS-DR12 galaxy clustering data, combined with {\it Planck} legacy cosmic microwave background (CMB) and baryon acoustic oscillation (BAO) measurements. We consider a set of models parameterized within the effective theory of structure formation (ETHOS), quantifying the
lifting of the $S_8$ tension in view of KiDS weak-lensing results. The most favorable scenarios point to a fraction $f\sim 10-100\%$ of interacting dark matter as well as a dark radiation temperature that is smaller by a factor $\xi\sim 0.1-0.15$ compared to the CMB, leading to a reduction of the tension to the $\sim 1\sigma$ level.
The temperature dependence of the interaction rate favored by relaxing the $S_8$ tension is realized for a weakly coupled unbroken non-Abelian $SU(N)$ gauge interaction in the dark sector.
To map our results onto this $SU(N)$ model, we compute higher-order corrections due to Debye screening. We find
a lower bound $\alpha_d\equiv g_d^2/(4\pi)\gtrsim 10^{-8} (10^{-9})$ for dark matter mass $1000 (1)$\,GeV for relaxing the $S_8$ tension, consistent with upper bounds from galaxy ellipticities and compatible with self-interactions relevant for small-scale structure formation.}
\maketitle
\newpage

\section{Introduction}
\label{sec:intro}

Our knowledge about the constituents, structure and evolution of the Universe has increased considerably by combining precision data covering a broad range of length and time scales. Although the $\Lambda$CDM model still agrees with many datasets, precise measurements have unveiled tensions in the values of cosmological parameters estimated by distinct probes, possibly indicating unaccounted systematic effects or alternatively hints for deviations from $\Lambda$CDM. First, the Hubble constant $H_0$ determined from the local distance ladder~\cite{Riess:2020fzl} and inferred indirectly within $\Lambda$CDM from the cosmic microwave background (CMB) anisotropies~\cite{Planck:2018vyg} as well as the scale of baryon acoustic oscillations (BAO)~\cite{eBOSS:2020yzd} disagree at a level of around 4$\sigma$ (see \cite{Verde:2019ivm,Riess:2019qba} for reviews and \cite{Schoneberg:2021qvd, DiValentino:2021izs} for summaries of proposed solutions). Second, the value of $S_8=\sigma_8(\Omega_m/0.3)^{0.5}$, with $\sigma_8$ being the amplitude of matter clustering on a scale of $8\,$Mpc$/h$ and $\Omega_m$ the matter density parameter, measured from a variety of large-scale structure (LSS) probes (including weak lensing~\cite{HSC:2018mrq,Heymans:2020gsg,KiDS:2020suj,Burger:2022lwh}, galaxy cluster counts~\cite{SPT:2018njh,Chiu:2022qgb}, and lensing combined with galaxy clustering~\cite{DES:2021wwk}) are systematically below the value inferred from CMB measurements within $\Lambda$CDM, with a significance of up to 3$\sigma$ (see also~\cite{Nunes:2021ipq,Amon:2022azi}). Third, the so-called small-scale crisis poses an intriguing puzzle~\cite{Bullock:2017xww,Hui:2016ltb}.

The aforementioned tensions can serve as a guideline to address another long-standing open question in cosmology: the nature of dark matter (DM). For example, the small-scale puzzles can be addressed in ultra-light~\cite{Ferreira:2020fam} or self-interacting~\cite{Tulin:2017ara} DM models. On the other hand, the $S_8$ tension points to scenarios where the matter power spectrum on scales $k\gtrsim 0.1 h/$Mpc is somewhat suppressed compared to larger scales of order $k\sim 10^{-2} h/$Mpc probed via CMB anisotropies. A priori, such a suppression can naturally arise in scenarios that deviate from the minimal paradigm of cold and collisionless DM. Examples include a non-zero free-streaming length for warm dark matter~\cite{Viel:2005qj}, a macroscopic de-Broglie wavelength such as for fuzzy dark matter~\cite{Hu:2000ke}, dark matter decay into invisible daughter particles~\cite{Simon:2022ftd}, or certain self-interacting dark matter scenarios~\cite{Egana-Ugrinovic:2021gnu}. Nevertheless, specific models imply particular patterns imprinted on the CMB and matter power spectra along with possible changes in the background evolution, such that it is non-trivial to identify viable and well-motivated models successfully addressing the $S_8$ tension.

A mechanism for suppressing power that is known to operate in nature is the Compton scattering of baryons (i.e.\ visible matter) with CMB photons in the early Universe. It is a plausible possibility that an analogous mechanism is responsible for a power suppression in (part of) the DM component. For example, DM can be gauged under a symmetry group akin to the SM. If the symmetry is unbroken and the gauge coupling sufficiently small, the gauge fields are light or massless, and we end up with a dark sector with two components~\cite{Lesgourgues:2015wza}: an \it interacting dark matter \rm (IDM) coupled with a new component of ultrarelativistic \it dark radiation \rm (DR). 

The scenario of DM-DR interactions can be realized microscopically by a variety of setups, leading in general to different implications for the power spectrum and structure formation.
On large scales, the effects of DM-DR models on structure formation can be parameterized through the effective theory of structure formation (ETHOS) \cite{Cyr-Racine:2015ihg}. ETHOS provides a (Boltzmann) framework to follow the evolution of DM and DR perturbations in presence of DM-DR interactions as well as DR self-interactions. DM-DR models have often been discussed in the context of small-scale puzzles (see e.g. \cite{Vogelsberger:2015gpr}) and also considered with respect to the
$H_0$ and $S_8$ tensions~\cite{Buen-Abad:2017gxg,Hooper:2022byl}. As far as the $H_0$ tension is concerned, they behave similar to models with extra radiation, meaning that they
cannot solve the $H_0$ tension in a minimal setup (see however~\cite{Aloni:2021eaq} using a two-component DR model and~\cite{Niedermann:2021vgd} assuming a phase transition
in the DR sector that addresses $H_0$).
However, the interaction of DM with DR allows to reduce $S_8$ relative to $\Lambda$CDM and therefore provides a possibility to address the $S_8$ tension~\cite{Buen-Abad:2017gxg}.

In this work we update constraints on DM-DR models within the ETHOS framework considering a range of scenarios depending on $(i)$ the temperature dependence of the interaction rate,  $(ii)$ the fraction of DM that is interacting, and $(iii)$ the case of free-streaming or efficiently self-interacting DR (also known as {\it dark fluid}). The main new ingredient of our analysis are the monopole, quadrupole and hexadecapole galaxy clustering power spectra in redshift space provided by DR12 of the BOSS galaxy survey~\cite{BOSS:2016wmc, Reid:2015gra}, that we combine with CMB \it Planck 2018 \rm legacy data \cite{Planck:2018vyg,Planck:2018lbu} and complementary BAO-scale measurements. We use the CLASS-PT framework~\cite{Chudaykin:2020aoj} to take the full shape information of the BOSS galaxy clustering data into account~\cite{Semenaite:2021aen,Sanchez:2013tga}. This implementation is based on a perturbative approach to non-linear corrections of the power spectrum of biased tracers in redshift space~\cite{DAmico:2019fhj,Ivanov:2019hqk}, built on top of the large-scale bias expansion (see~\cite{Desjacques2016} for a review) and an effective field theory description of redshift space clustering~\cite{Baumann:2010tm,Foreman:2015lca,Perko:2016puo,Konstandin:2019bay}.
We analyze the capability of the various scenarios to address the $S_8$ tension, comparing also to a prior obtained from KiDS-1000~\cite{KiDS:2020suj}.

The features of the DM-DR interaction that are most promising to solve the $S_8$ tension can be realized by a $SU(N)$ gauge symmetry within the dark sector, leading to a suitable temperature dependence of the interaction rate as well as DR self-interaction~\cite{Lesgourgues:2015wza}. Both of these properties are intimately related to the non-Abelian gauge boson interaction. We therefore interpret our results within the parameters of this model. The interaction rate is sensitive to Debye screening and we compute and include higher order corrections in the mapping between ETHOS and the microscopic parameters of the theory.

The structure of this work is as follows. In Sec.~\ref{sec:theory} we review the theoretical basis of both the ETHOS framework as well as the perturbative description of the galaxy power spectrum in redshift space within the effective theory approach. Next, in Sec.~\ref{sec:data} we present the datasets and statistics used in this work and detail the Markov chain Monte Carlo (MCMC) setup. The main results are presented in Sec.~\ref{sec:results}. Finally, we discuss in Sec.~\ref{sec:mapping} the mapping between ETHOS and a dark sector described by a $SU(N)$ gauge theory. We conclude in Sec.~\ref{sec:conslusion}. We dedicate appendices to complementary material of IDM-DR modeling and a complete calculation of the DM-DR drag opacity for the $SU(N)$ model, taking the complete leading-order as well as higher-order corrections into account that have been neglected so far.

\section{Prerequisites}
\label{sec:theory}

In this section we review the theoretical basis of this work. We start with the ETHOS formulation of DM-DR interaction. Then we summarize the large-scale bias expansion and the perturbative description of the power spectrum in redshift space within an effective field theory approach. 

\subsection{The ETHOS parameterization} \label{sec:ETHOS}

ETHOS considers a non-relativistic DM species coupled to a DR component, taking the coupled hierarchy of moments of the phase-space distribution function for DR into account. The evolution equations for the energy density contrast $\delta$, velocity divergence $\theta$ and higher moments $\Pi_l$ for DR are given by~\cite{Cyr-Racine:2015ihg}
\begin{eqnarray}
\dot{\delta}_{\rm DR} + \frac{4}{3}\theta_{\rm DR} - 4 \dot{\phi}_g &=& 0 \,,
\\
\dot{\theta}_{\rm DR} + k^2\left( \sigma^2_{\rm DR} - \frac{1}{4}\delta_{\rm DR}\right) - k^2\psi_g &=& \Gamma_{\rm DR-IDM}\left( \theta_{\rm DR} - \theta_{\rm IDM}\right)  \,,
\\
\dot{\Pi}_{\rm DR,l} + \frac{k}{2l+1}\left[ (l+1) \Pi_{\rm DR,l+1} - l \Pi_{\rm DR,l-1} \right] &=& \left( \alpha_l \, \Gamma_{\rm DR-IDM} + \beta_l \, \Gamma_{\rm DR-DR}\right)\Pi_{\rm DR,l}\,, \label{eq:stress_DR}
\end{eqnarray}
where $\{\alpha_l,\, \beta_l\}$ are the ($l$-dependent) angular coefficients for IDM-DR and DR-DR scatterings, respectively. $\phi_g$ and $\psi_g$ are the gravitational potentials and $\sigma_{\rm DR} = \Pi_{\rm DR,2}/2$ is the shear stress.  The opacity of each process is given by $\Gamma$ (sometimes also named $\dot{\kappa}$). 

For non-relativistic IDM only the density contrast and velocity divergence need to be taken into account,
\begin{eqnarray}\label{eq:idm_evolve}
\dot{\delta}_{\rm IDM} + \theta_{\rm IDM} - 3 \dot{\phi}_g &=& 0 \,,
\\
\dot{\theta}_{\rm IDM} - c_{\rm IDM}^2k^2\delta_{\rm IDM} + \mathcal{H}\theta_{\rm IDM} - k^2\psi_g &=& \Gamma_{\rm IDM-DR}\left( \theta_{\rm IDM} - \theta_{\rm DR}\right) \,,\label{eq:idm_evolve2}
\end{eqnarray}
where $\mathcal{H}=aH$ is the Hubble parameter rescaled by the scale factor and $c_{\rm IDM}$ is the IDM sound velocity, which is small for non-relativistic IDM and does not contribute on large scales. Non-linear corrections to Eqs.~(\ref{eq:idm_evolve}) and (\ref{eq:idm_evolve2}) are included as presented in Sec.~\ref{sec:bias_expansion}.
 We can expand the opacities $\Gamma$ as a power-law in temperature and therefore in redshift,
\begin{eqnarray}\label{eq:gamma_dr_dm}
\Gamma_{\rm DR-IDM} &=& - (\Omega_{\rm IDM}h^2)\,x_{\rm IDM}(z)\sum_n a_n \left( \frac{1+z}{1+z_{\rm D}}\right)^n\,, 
\\ \label{eq:gamma_dr_dr}
\Gamma_{\rm DR-DR} &=& - (\Omega_{\rm DR}h^2)\,x_{\rm DR-DR}(z)\sum_n b_n \left( \frac{1+z}{1+z_{\rm D}}\right)^n\,,
\end{eqnarray}
and using energy-momentum conservation we can relate $\Gamma_{\rm IDM-DR}$ and  $\Gamma_{\rm DR-IDM}$,
\begin{equation} \label{eq:gamma_dm_dr}
\Gamma_{\rm IDM-DR} = \frac{4}{3}\left( \frac{\rho_{\rm DR}}{\rho_{\rm IDM}}\right)\Gamma_{\rm DR-IDM}   = - \frac{4}{3}(\Omega_{\rm DR}h^2)\,x_{\rm IDM}(z)\sum_n a_n  \frac{(1+z)^{n+1}}{(1+z_{\rm D})^n}\,.
\end{equation}
The dimensionless functions $x_{\rm IDM}(z)$ and $x_{\rm DR-DR}(z)$ absorb non-trivial redshift dependences, i.e.~deviations from a power-law scaling, and are for simplicity set to unity in this work. 
Furthermore, we use $z_{\rm D} = 10^7$ as a normalization and truncate the series in the first (dominant) term in $n$. 

The collision terms are then encapsulated in the set of parameters $\{ \alpha_l ,\, \beta_l,\, a_n,\, b_n\}$, such that distinct microscopic models can be mapped onto those coefficients (see Sec.~\ref{sec:mapping} for an example). Regarding DR, we consider two limiting cases: 
\begin{itemize}
\item strongly self-interacting DR (also known as \emph{dark fluid}) that leads to an efficient damping of higher moments $\Pi_{\rm DR,l}\to 0$ for all $l\geq 2$ (formally corresponding to $\Gamma_{\rm DR-DR} \rightarrow \infty$), such that the parameters $\alpha_l,\beta_l,b_n$ are irrelevant,
\item and DR that does not interact with itself at all ($\Gamma_{\rm DR-DR} \rightarrow 0$, i.e. $b_n=0$), in which case we set $\alpha_l = 1$ while $\beta_l$ is irrelevant. Since there is no self-interaction, we follow the usual convention to refer to this case as  {\it free-streaming}, though DR still interacts with IDM. 
\end{itemize}
The most relevant parameters are the amplitude of DM-DR interaction and its temperature dependence. We parameterize the former by
\be
  a_{\rm dark} \equiv a_n\,,
\ee 
and consider $n=0,2,4$. The case $n=0$ corresponds to the $SU(N)$ interaction discussed in Sec.\,\ref{sec:mapping}, while $n=2$ and $n=4$ arise for example for an unbroken Abelian $U(1)$ interaction or a massive vector mediator, respectively~\cite{Cyr-Racine:2015ihg}. In addition, there is one parameter characterizing the density or equivalently temperature of dark radiation. We adopt
\begin{equation}
\xi \equiv \frac{T_{\rm DR}}{T_{\rm CMB}}\biggl|_{z=0}\,,
\end{equation}
written in terms of the ratio between the dark radiation temperature $T_{\rm DR}$ and the CMB temperature $T_{\rm CMB}$. This quantity can also be expressed in terms of extra light species as $\Delta N_{\rm eff} \equiv \rho_{\rm DR}/\rho_{1\nu}$  (also called $\Delta N_{\rm fluid}$ for the dark fluid case), where $\rho_{1\nu}$ is the energy density of one neutrino flavour. See App.~\ref{app:xi_delta_mapping} for more details. The density parameter of DR is given by
\begin{equation} \label{eq:omega_DR}
\omega_{\rm DR} = \Omega_{\rm DR} h^2 = \frac{\eta_{\rm DR}}{2}\zeta\, \xi^4\, \Omega_\gamma h^2\,,
\end{equation}
where $\zeta = 1\,(7/8)$ for bosons (fermions) and $\eta_{\rm DR}$ is the DR spin/color degeneracy factor. For concreteness, we take DR to be bosonic and with $\eta_{\rm DR} = 2$ unless stated otherwise, but note that our results can be rescaled to other values while keeping $\omega_\text{DR}$ fixed (see Sec.\,\ref{sec:mapping}). Finally, we consider the possibility that only a fraction 
\begin{equation}
f \equiv \frac{\Omega_{\rm IDM}}{\Omega_{\rm CDM} + \Omega_{\rm IDM}}\,,
\end{equation}
of the total DM interacts with DR, with the remaining fraction behaving as usual cold dark matter (CDM).

Typically, two interesting limits of those parameters are discussed in the literature \cite{Buen-Abad:2017gxg}. First, 
the \it weakly interacting \rm (WI) limit \cite{Buen-Abad:2015ova,Lesgourgues:2015wza}, in which the DM-DR interaction is comparable to or smaller than the Hubble rate around matter-radiation equality ($\Gamma_{\rm IDM-DR} (a_{\rm eq}) \lesssim \mathcal{H}(a_{\rm eq})$). Since the power suppression for that case is milder, this kind of model in general \it allows \rm for $f = 1$. Second, the \it strongly coupled limit \rm for which DM and DR form a tightly coupled \it dark plasma \rm (DP) \cite{Buen-Abad:2017gxg, Chacko:2016kgg}, in which $\Gamma_{\rm IDM-DR} (a_{\rm eq}) \gg {\cal H}(a_{\rm eq})$. The precise value of $\Gamma_{\rm IDM-DR}\propto a_\text{dark}$ drops out in the tight coupling limit, but this case is typically only viable when \it demanding \rm $f \ll 1$. We will discuss those scenarios further below. 

Finally, we note that an interesting feature of the case $n=0$ is that the ratio $\Gamma_{\rm IDM-DR}/{\cal H}$ has only a weak time dependence and therefore the interaction can be active while a wide range of scales enters the horizon, leading to a relatively mild scale dependence of the power suppression~\cite{Buen-Abad:2015ova} (see e.g. left panel of Fig. 3 in~\cite{Buen-Abad:2017gxg}). The amount of DR, given by the parameter $\xi$, mainly determines the scale where the suppression in the power spectrum sets in, while the interaction strength $a_{\rm dark}$ and the fraction $f$ determine the amount of suppression. This feature of $n=0$ is favorable in view of the $S_8$ tension, and our results confirm this observation as we shall see below (see also Appendix~\ref{app:idm_extra}). 

\subsection{The large-scale bias expansion} \label{sec:bias_expansion}

To relate the observed galaxy density contrast $\delta_g$ to the underlying matter distribution we use the large-scale bias expansion, in which we write the tracer field (in our case the galaxy overdensity) in terms of the most general set of independent operators $\mathcal{O}$~\cite{1984ApJ...284L...9K,Chan:2012jj,Assassi2014}
\begin{equation}
\label{eqn: bias_expansion}
    \delta^{g}(\boldsymbol{x}, z)=\sum_{\mathcal{O}} b_{\mathcal{O}}(z) \mathcal{O}(\boldsymbol{x}, z) + \epsilon(\boldsymbol{x}, z) + \sum_\mathcal{O}\epsilon_\mathcal{O}(\boldsymbol{x}, z)\mathcal{O}(\boldsymbol{x}, z)\,.
\end{equation}
Each operator is accompanied by its respective bias parameter $b_{\mathcal{O}}$.\footnote{Though each bias parameter has a very clear physical interpretation (see \cite{Desjacques:2016bnm} for a comprehensive review) and many efforts are spent on improving theoretical constraints and relations among them \cite{Barreira:2021ueb,Lazeyras:2021dar,Barreira:2021ukk}, we treat them here as free nuisance parameters to be include the in the MCMC runs.} The $\epsilon$ fields take into account the stochasticity of structure formation. The set of operators, up to third-order in perturbation theory\footnote{When counting each gradient and each power of a linear field as contributing one order.} is given by
\begin{equation}
\label{eqn: operator_basis}
    \mathcal{O} \in
    \left\{
    \delta,\, \delta^{2},\,  \delta^{3},\,  \mathcal{G}_{2}[\phi_{g}],\,  \delta\mathcal{G}_{2}[\phi_{g}],\,  \mathcal{G}_{3}[\phi_{g}],\,  \Gamma_{3}[\phi_{g}, \phi_{v}],\,  \nabla^2 \delta \right\}\,,
\end{equation}
where $\phi_{v}$ is velocity potential and $\phi_{g}$ the gravitational potential. $\mathcal{G}_n$ are the $n$-order Galileon operators and $\Gamma_3$ is the difference between the Galileon for the density and of the velocity potentials \cite{Chan:2012jj,Assassi2014}. 
The coefficient of the last operator can be viewed as a nonlocal bias and in addition absorbs small-scale uncertainties within an effective field theory treatment of dark matter clustering itself \cite{Baumann:2010tm, Carrasco:2012cv}.

Taking the correlations among all pairs of operators into account, we can write the galaxy-galaxy power spectrum as
\begin{eqnarray} 
\label{eq:generalP}
 P^{gg}(z,k) &=& \left(b_1\right)^2 \left[ P_{\rm lin}(z,k) + P_{\rm 1L}(z,k) \right] + b_1b_2\,\mathcal{I}_{\delta^2}(z,k) + 2b_1b_{\mathcal{G}_2}\,\mathcal{I}_{\mathcal{G}_2}(z,k) \nonumber \\
 &+& \left(2b_1b_{\mathcal{G}_2} + \frac{4}{5} b_1b_{\Gamma _{3}}\right)\mathcal{F}_{\mathcal{G}_2}(z,k)  
+ \frac{1}{4}\left(b_2\right)^2\,\mathcal{I}_{\delta^2\delta^2}(z,k)  \\ &+&\left(b_{\mathcal{G}_2}\right)^2\,\mathcal{I}_{\mathcal{G}_2\mathcal{G}_2}(z,k) + b_2b_{\mathcal{G}_2}\,\mathcal{I}_{\delta^2\mathcal{G}_2}(z,k) + P_{\nabla^2\delta}(z,k) + P_{\varepsilon\varepsilon}(z,k)  \,, \nonumber
\end{eqnarray}
where $P_{\rm lin}$ is the linear matter power spectrum and $P_{\rm 1L}$ the one-loop contribution. To avoid cluttering the text, we have omited the $z$ dependence of all bias parameters. For a full expression for $\mathcal{I}_{\delta^2}, \mathcal{I}_{\mathcal{G}_2}\,, \mathcal{F}_{\mathcal{G}_2}\,,\mathcal{I}_{\delta^2\delta^2}\, \mathcal{I}_{\mathcal{G}_2\mathcal{G}_2}\, \textrm{and} \, \mathcal{I}_{\delta^2\mathcal{G}_2}$, see \cite{Chudaykin:2020aoj}. For a generalization of the large-scale bias expansion to more tracers, see \cite{Mergulhao:2021kip}. The two last terms correspond respectively to a term proportional to $\langle\nabla^2\delta\delta\rangle\sim k^2P_\text{lin}$ and the stochastic contribution proportional to $k^0$ and $k^2$.

In redshift space, the most general expression for the galaxy-galaxy power spectrum is constructed using a multipole expansion with respect to the angle between the Fourier wavevector and the line-of-sight direction $\hat{\mbf{z}}$. For the full expression for the monopole, quadrupole and hexadecapole used in this work, we refer again to \cite{Chudaykin:2020aoj}. We use the same bias parameters and structure as described in~\cite{Philcox:2021kcw}. We use CLASS-PT~\cite{Chudaykin:2020aoj} to calculate each term in the power spectrum. CLASS-PT uses the FFT-Log algorithm from \cite{Schmittfull:2016jsw,Simonovic:2017mhp} to boost the integral calculations. It also takes into account the Alcock-Paczynski (AP) distortions~\cite{AP} from assuming a fiducial cosmology for converting angles and redshift differences in distances transverse and along the line of sight, respectively, and implements the IR-resummation~\cite{Eisenstein:2006nj} from~\cite{Blas:2016sfa}.

\section{Datasets and statistics} \label{sec:data}

We start this section by exposing the different datasets considered in this work. Next, we explain the statistics used to compare how the models perform compared to $\Lambda$CDM and to quantify the $S_8$ tension in those models. 

\subsection{Datasets} \label{dataset}
For this work, we consider the following datasets:
\begin{itemize}

\item \bf Planck 2018\rm: We combine {\it Planck 2018} high-$\ell$ TT+TE+EE, low-$\ell$ TT+EE and lensing data \cite{Planck:2018vyg,Planck:2018lbu}. 

\item \bf BAO(+RSD)\rm: BAO data is added from the 6dFGS measurements at $z=0.106$ \cite{Beutler:2011hx}, SDSS at $z=0.15$ \cite{Ross:2014qpa} and from SDSS-III DR12 \cite{BOSS:2016wmc}. Moreover, we include eBOSS DR14 Lyman-$\alpha$ auto-correlation \cite{deSainteAgathe:2019voe} and cross-correlation with quasars \cite{Blomqvist:2019rah} (see also \cite{Cuceu:2019for}). 
When \emph{not} including FS in the analysis, we also include the measurement of $f\sigma_8$ from redshift space distortion (RSD) of BOSS DR12 \cite{BOSS:2016wmc}. 

\item \bf KiDS\rm: Simillarly to \cite{Simon:2022ftd}, we implement the KiDS-1000 likelihood from \cite{KiDS:2020suj} as a split-normal likelihood with $S_8 = 0.759\substack{+0.024\\-0.021}$.

\item \bf Full-shape (FS)\rm: Data from BOSS DR12 from different cuts of the sky (NGC and SGC) \cite{BOSS:2016wmc, Reid:2015gra, Kitaura:2015uqa}, divided  in two $z$ slices between $0.2<z<0.5$ ($z_1$) and $0.5<z<0.75$ ($z_3$) . The likelihoods\footnote{Available at \url{https://github.com/oliverphilcox/full_shape_likelihoods}.} are constructed using the covariance matrix estimated from a set of 2048 mock catalogs from ‘MultiDark-Patchy’ \cite{Abbetal,Rodriguez-Torres:2015vqa} and the window(free) function of \cite{Philcox:2020vbm,Philcox:2021ukg}. We use the information from the power spectrum monopole, quadrupole and hexadecapole (up to $k_{\rm max} = 0.2h\,\textrm{Mpc}^{-1}$), real-space power spectrum proxy from \cite{Scoccimarro:2004tg,Ivanov:2021fbu} (up to $k_{\rm max} = 0.4h\,\textrm{Mpc}^{-1}$)\footnote{This refers to a combination of redshift-space multipoles that approximate the power spectrum perpendicular to the line-of-sight~\cite{Scoccimarro:2004tg} (termed $Q_0$ in~\cite{Ivanov:2021fbu}), reducing the impact of the fingers-of-god effect, and therefore allowing for an extension to smaller scales ($k_{\rm max} = 0.4h/\textrm{Mpc}^{-1}$). We take into account the cross-covariance between $Q_0$ data and the individual multipoles in order to avoid double counting~\cite{Philcox:2021kcw}.}, the BAO post-reconstructed spectrum from \cite{Philcox:2020vvt}, similarly to what was done in~\cite{Philcox:2021kcw} and based on~\cite{BOSS:2016hvq}. For the priors in the bias and counter-term parameters from the large-scale bias expansion of Sec.~\ref{sec:bias_expansion} we considered the same setup as described in \cite{Philcox:2021kcw}, with each bias and stochastic parameters being fit separately in each dataset. We explicitly checked that none of the bias terms reaches the prior bounds.\footnote{A recent analysis of the full shape pipeline pointed out the impact of prior volume effects on the posterior of cosmological parameters when used with BOSS data and without including {\it Planck} data~\cite{Simon:2022lde}. Here we use FS information only in combination with Planck. Furthermore, we use statistical indicators based on best-fit values (see below) that are by construction not affected by volume effects. Finally, in light of some possible discrepancy of BAO reconstructed data~\cite{Gil-Marin:2015nqa,BOSS:2016hvq} also pointed out by~\cite{Simon:2022lde}, we checked that our results are not affected by including/removing the contribution from reconstructed BAO data.} 

\end{itemize}

Akin done by \cite{Simon:2022ftd}, we always include {\it Planck} and BAO data in our analysis. We focus on comparing three main setups: 
\begin{itemize}
\item {\it Planck} + BAO + RSD\footnote{When FS is incuded, we do not include RSD from BOSS.}, 
\item {\it Planck} + BAO + FS and 
\item {\it Planck} + BAO + FS + KiDS
\end{itemize}
The inclusion of KiDS in the analysis has \it not \rm the intend to provide a full joint analysis of KiDS with other datasets, since that would require a model-specific KiDS likelihood. 
Instead, including the KiDS prior serves as an indication of how well the IDM-DR models can account for the low-$S_8$ trend seen by KiDS. Moreover, combining datasets that disagree is, of course, troublesome, e.g. for the case of {\it Planck} and KiDS for $\Lambda$CDM. As we will point out, this is \it not \rm the case for KiDS and {\it Planck} + BAO + FS for interacting DM.  

Regarding the cosmological parameters, we explore the following parameter space: 
\begin{equation}
\{\omega_b,\,   \omega_{\rm cdm},\,  \log{\left(10^{10}A_s\right)},\, n_s, \, \tau_{\rm reio},\, H_0,\, \xi,\, a_{\rm dark} \}\,.
\end{equation}
We consider as our benchmark \it fiducial \rm scenario the \it dark fluid \rm limit $\Gamma_{\rm DR-DR} \rightarrow \infty$ for DR self-interaction, with $f=1$ and $n=0$. Variations of the IDM fraction $f$ are discussed in Sec.~\ref{sec:fchange}. We also consider in Sec.~\ref{sec:freestreaming} the \it free-streaming \rm scenario $\Gamma_{\rm DR-DR} \rightarrow 0$ of DR without self-interaction, and considering $n=0,2,4$.  

We have used MontePython \cite{Audren:2012vy, Brinckmann:2018cvx} to run the Markov chain Monte Carlo. MontePython connects to CLASS \cite{Blas:2011rf} to solve the Boltzmann system of equations. As previously mentioned, the non-linear power spectra were calculated using CLASS-PT.\footnote{We note though that CLASS-PT was implemented on top of CLASS v.2.6, while the ETHOS parameterization was implemented only in CLASS v.2.9 \cite{Archidiacono:2017slj}. We, therefore, have modified CLASS-PT to include the ETHOS model.} We used the Gelman-Rubin criteria \cite{GR} for the convergence of the chains ($R-1<0.02$). For $a_{\rm dark}$ we sampled the chains in log coordinates, using the prior $\log_{10} \left(a_{\rm dark}\right) < 20$, similarly to \cite{Archidiacono:2019wdp}.\footnote{Note that the IDM model also suffers from volume effects when $\xi \rightarrow0$. A more in-depth investigation of those volumes effects using e.g.\ the profile likelihood (see e.g.~\cite{Herold:2021ksg,Gomez-Valent:2022hkb}) would still be interesting. In our case, as discussed later in the text, those volume effects are not relevant since one of the main results we point out shows that $\xi$ deviates from zero when including the BOSS likelihood (when considering the low $S_8$ region and/or including KiDS). In addition, for model comparison we use statistical indicators that are based on frequentist statistics (see Sec.~\ref{sec:Statistics} below) and are therefore by construction independent of volume effects.} Note that an upper bound for $a_{\rm dark}$ is needed, otherwise for $\xi\rightarrow0$, an arbitrarily large value of $a_{\rm dark}$ is allowed. For $\xi$ we set a conservative prior $\xi < 1$, inside the CMB bounds for $N_{\rm eff}$.

One last comment is in order. Within IDM-DR models, the interaction between both components is only relevant during the early stages of our Universe. This means that for the latest stages of evolution, IDM behaves as CDM on large scales with its interaction with DR being strongly suppressed. Therefore the perturbative computation of power spectra based on perfect fluid equations complemented with an effective stress tensor is still valid and the interactions do not have to be modeled at the latest (non-linear) stages of cosmic expansion. We refer the reader to Appendix~\ref{app:n_scaling} for more details. 

\subsection{Statistics}\label{sec:Statistics}
In order to quantify how well a model performs in fitting the data compared to $\Lambda$CDM, we use the $\chi^2$ difference.
The $\chi^2$ difference for a model $\mathcal{M}$ with respect to $\Lambda$CDM is given by
\begin{equation}
\Delta  \chi^2_{\,\mathcal{M},\rm data} = \chi^2_{\rm min, \mathcal{M},\rm data} - \chi^2_{\rm min, \Lambda CDM, data} \,,
\end{equation}
where $\chi^2_{\rm min}$ is the best-fit value.\footnote{Note that the $\Delta  \chi^2$ is very dependent on a single value in the posterior, which is the best-fit point. Finding the global minima in a MCMC run is not always a trivial task depending on the topology of the likelihood. For some cases, it was necessary to perform multiple independent runs, each with six chains and $\mathcal{O}(10^6)$ points.} From $\Delta  \chi^2$ we can define the Akaike Information Criterium (AIC)
\begin{equation}
\Delta  {\rm AIC}_{\,\mathcal{M},\rm data} = \Delta  \chi^2_{\,\mathcal{M},\rm data}  + 2(N_\mathcal{M} - N_{\rm \Lambda CDM})\,,
\end{equation}
where $N$ is the number of degrees of freedom of the model. The AIC is a way to penalize models that have too many free parameters using Occam’s razor criteria.

In order to quantify a tension in a dataset within a model (and therefore with the same number of free parameters), we use the `difference in maximum a posterior' as \cite{Raveri:2018wln}
\begin{equation}
Q_{\rm DMAP}^{\mathcal{M},\rm data} = \chi^2_{\rm min,\mathcal{M}}({\rm w/} \,\, {\rm data}) - \chi^2_{\rm min,\mathcal{M}}({\rm w/o}\,\, {\rm data})\,.
\end{equation}
We can quantify how much a dataset is in tension with other datasets by taking the square root of $Q_{\rm DMAP}^{\mathcal{M},\rm data}$. The tension of a model $\mathcal{M}$ with KiDS can be quantified by
\begin{equation}
 {\rm Tension} = \sqrt{Q_{\rm DMAP}^{\mathcal{M},\rm KiDS}} = \sqrt{ \chi^2_{\rm min,\mathcal{M}}({\rm w/} \,\, {\rm KiDS}) - \chi^2_{\rm min,\mathcal{M}}({\rm w/o}\,\, {\rm KiDS})}\,.
\end{equation}
Notice that the statistics used here to quantify both the tension with respect to KiDS data and the preference of a model over $\Lambda$CDM resemble the statistics used to discuss how different models address the $H_0$ tension with respect to SH0ES in \cite{Schoneberg:2021qvd}.

\section{Results}
\label{sec:results}

In this section we discuss results for different models within the ETHOS framework. We consider three different scenarios, already pointed out in Sec.~\ref{dataset}: {\it Planck} + BAO +RSD, {\it Planck} + BAO + FS and {\it Planck} + BAO + FS + KiDS. We stress again that the posteriors when including KiDS should be taken with a grain of salt: the KiDS result is implemented as a Gaussian around $S_8$ measured for $\Lambda$CDM and not as a full model-dependent likelihood. Here we show its effect for comparison with other models and to figure out directions in parameter space that are preferred when the low $S_8$ value reported by KiDS is included. 

We start in Sec.~\ref{sec:fiducial} discussing the fiducial case, in which all of DM interacts with DR and the latter behaves as a dark fluid. Then, we dedicate Sec.~\ref{sec:fchange} for studying scenarios with different IDM fractions and Sec.~\ref{sec:freestreaming} for the case of free-streaming DR. The main results are summarized in Tab.\,\ref{tab:resultstension}.

\begin{table}[ht]
	\footnotesize
	\begin{center}
		\begin{tabular}{ | l || c | c |  c || c | c | c || c |c|c||c|}
			\hline
			\multirow{2}{*}{} & 
			\multicolumn{3}{|c||}{ {\it Planck} + BAO} &
			\multicolumn{3}{|c||}{ + FS} & 
			\multicolumn{3}{|c||}{ + FS + KiDS} & \multirow{2}{*}{{\scriptsize$\sqrt{Q_{\rm DMAP}^{\rm KiDS}}$}}\\ 
			\cline{2-10}
			&$\chi^2$&$\Delta \chi^2$&$\Delta$AIC&$\chi^2$&$\Delta \chi^2$&$\Delta$AIC&$\chi^2$&$\Delta \chi^2$&$\Delta$AIC& \\ \hline
			
			$\Lambda$CDM & \color{black}{2791.6} & - & - & \color{black}{3563.1} & - & - & \color{black}{3571.6} & - & -&\color{black}{2.9}$\sigma$
			\\ \hline

			Fluid: &  &  &  &  &  &  &  &  & &
			\\ 
			
			\quad Fiducial  & \color{black}{2790.3} & \color{black}{-1.3} & \color{black}{2.7} & \color{black}{3557.7} & \color{black}{-5.5} & \color{black}{-1.5} & \color{black}{3559.9} & \color{black}{-11.7} & \color{black}{-7.7}&\color{black}{1.5}$\sigma$
			
			\\

			\quad $f = 0.1$  & \color{black}{2789.5} & \color{black}{-2.1} & \color{black}{1.9} & \color{black}{3560.4} & \color{black}{-2.7} & \color{black}{1.3} & \color{black}{3562.0} & \color{black}{-9.6} & \color{black}{-5.6}&\color{black}{1.3}$\sigma$
			\\

			\quad $f = 0.01$ & \color{black}{2791.1} & \color{black}{-0.5} & \color{black}{3.5} & \color{black}{3561.5} & \color{black}{-1.6} & \color{black}{2.4} & \color{black}{3566.9} & \color{black}{-4.7} & \color{black}{-0.7}&\color{black}{2.3}$\sigma$ \\

			\quad $f$ free & \color{black}{2790.3} & \color{black}{-1.3} & \color{black}{4.7} & \color{black}{3558.0} & \color{black}{-5.2} & \color{black}{0.8} & \color{black}{3561.4} & \color{black}{-10.2} & \color{black}{-4.2}&\color{black}{1.9}$\sigma$
			\\ \hline

			{\scriptsize Free-streaming:} &  &  &  &  &  &  &  &  & &
			\\ 
			
			\quad $n=0$ & \color{black}{2790.2} & \color{black}{-1.4} & \color{black}{2.6} & \color{black}{3559.0} & \color{black}{-4.1} & \color{black}{-0.1} & \color{black}{3559.7} & \color{black}{-11.9} & \color{black}{-7.9}&\color{black}{0.8}$\sigma$
			
			\\

			\quad $n=2$& \color{black}{2790.0} & \color{black}{-1.6} & \color{black}{2.4} & \color{black}{3560.7} & \color{black}{-2.4} & \color{black}{1.6} & \color{black}{3567.7} & \color{black}{-3.9} & \color{black}{0.1}&\color{black}{2.6}$\sigma$
			\\
			
			\quad $n=4$ & \color{black}{2790.7} & \color{black}{-0.9} & \color{black}{3.1} & \color{black}{3559.9} & \color{black}{-3.2} & \color{black}{0.8} & \color{black}{3568.0} & \color{black}{-3.6} & \color{black}{0.4}&\color{black}{2.8}$\sigma$  \\ 
			
			\hline

		\end{tabular}
	\end{center}
	\caption{\small Total $\chi^2$, $\Delta \chi^2$, $\Delta{\rm AIC}$ and $\sqrt{Q_{\rm DMAP}^{\rm KiDS}}$ for the best-fit of different models. We compare distinct dataset combinations (see Sec.~\ref{sec:Statistics}). When considering the fluid scenario, we fix $n=0$. The fiducial and free-streaming scenarios correspond to $f=1$. 
	}
	\label{tab:resultstension}
\end{table}

\subsection{Fiducial (dark fluid, $f=1$, $n=0$) scenario} \label{sec:fiducial}

We start by presenting the results for the fiducial scenario, in which DR is described as a dark fluid, all DM interacts with DR ($f=1$), and the temperature dependence of the interaction rate is comparable to that of the Hubble rate ($n=0$).

\begin{figure}[h]
	\centering  
	\includegraphics[width=.65\textwidth]{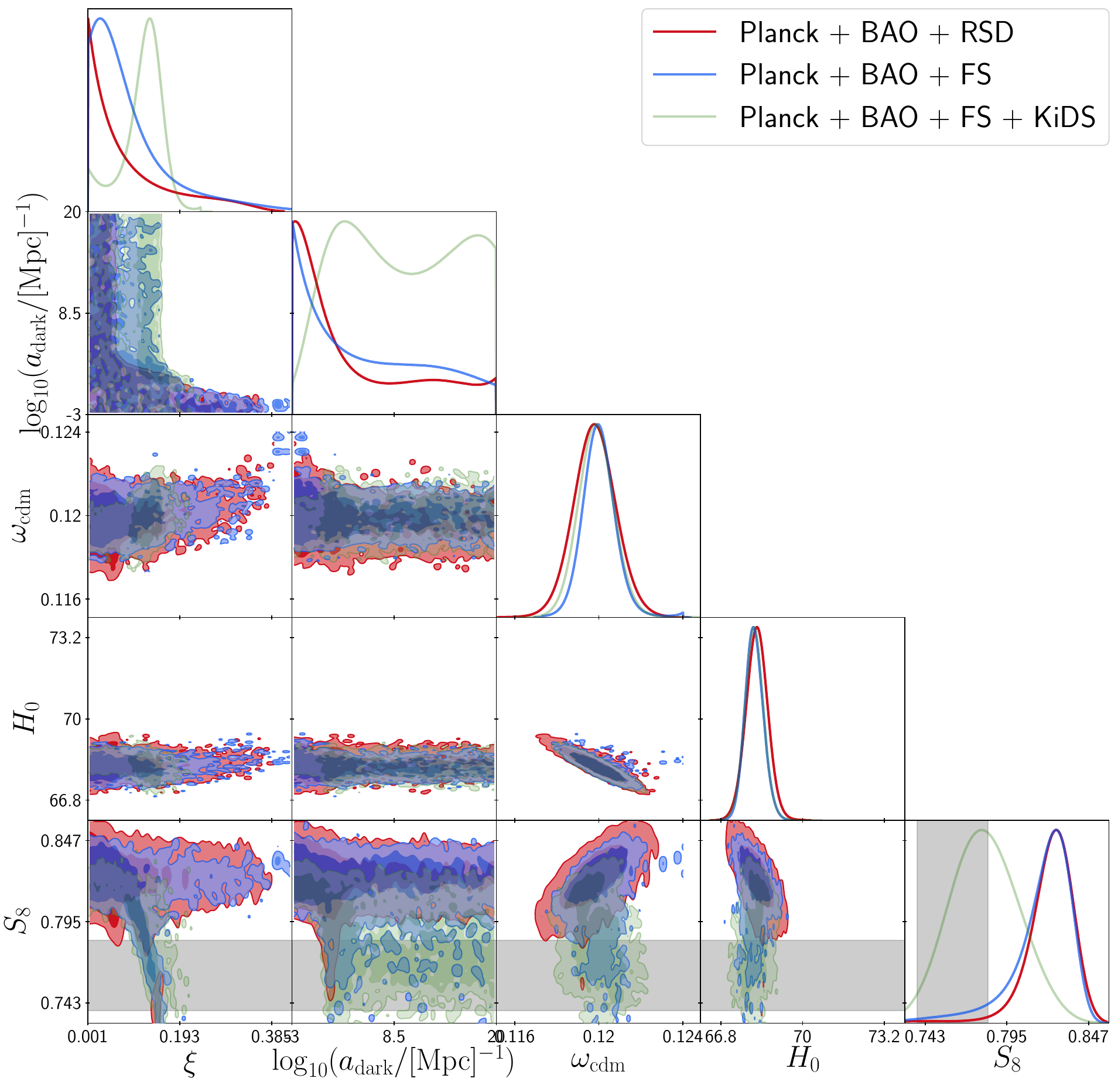}
	\includegraphics[width=.3\textwidth]{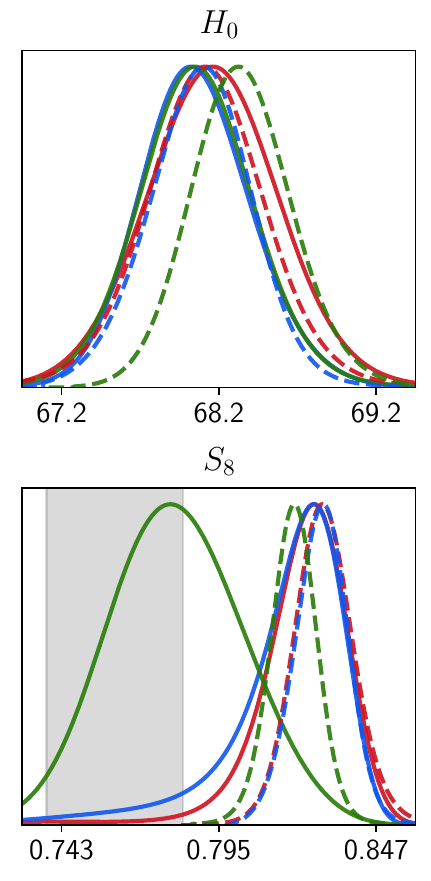}
	\caption{\label{fig:corner_lcdm_fid}
		\small Left: MCMC result for the fiducial scenario with $f=1$ for three different dataset combinations: {\it Planck} + BAO + RSD (red), {\it Planck} + BAO + FS (blue) and {\it Planck} + BAO + FS + KiDS (green) . The KiDS $1\sigma$ bounds on $S_8$ are shown as gray bands. Right: posteriors of $S_8$ and $H_0$ for the fiducial model (solid) and for $\Lambda$CDM (dashed) for the same dataset combinations.}
\end{figure}

In the right panel of Fig.~\ref{fig:corner_lcdm_fid}, we show the results for the posterior of $S_8$ for $\Lambda$CDM (dashed) and the fiducial model (solid). 
The difference between the red dashed line and the gray band, representing the nominal $1 \sigma$ KiDS bounds, clearly shows the tension between {\it Planck} + BAO and weak-lensing data within $\Lambda$CDM. Combining {\it Planck} and BAO with FS does not change the posterior to smaller $S_8$ values (blue dashed line), indicating that {\it Planck} data dominates the posterior.

For the fiducial IDM model, we see that {\it Planck} + BAO + RSD (red solid) already prefer smaller values of $S_8$ if compared to $\Lambda$CDM. Not only does the width of the posterior increase, which is expected from adding more free parameters, but also a skew in the direction of smaller $S_8$ arises (note that the same does not happen on the right side of the posterior). The inclusion of FS leads to an even more skewed tail and a small shift towards low $S_8$, compatible with \cite{Philcox:2021kcw}\footnote{Note that \cite{Zhang:2021yna} finds a larger value of $S_8$ when compared to \cite{Philcox:2021kcw}. We checked that our results for $\Lambda$CDM when including {\it Planck} are compatible with  \cite{Philcox:2021kcw}. See also \cite{Simon:2022lde} for a case without {\it Planck}.}. We see that {\it Planck}+BAO+FS is \it not \rm incompatible with KiDS for IDM and in that case it makes sense to combine both datasets. The $S_8$ tension is alleviated from $2.9\sigma$ for $\Lambda$CDM to $1.5\sigma$ for the fiducial IDM model (see Tab.\,~\ref{tab:resultstension}). In Sec.\,\ref{sec:fchange} we will see that models with a reduced fraction of IDM are more effective
in reducing the tension. Moreover, one may wonder how degenerate are the bias parameters of the FS analysis with a shift in $S_8$.
We explicitly checked that the shifts in $S_8$ are not being trivially absorbed by $b_1$ or other bias parameters.

In the left panel of Fig.~\ref{fig:corner_lcdm_fid} we display the combined posteriors for the fiducial IDM cosmology. We can see, e.g., in the $S_8$ vs.~$\xi$ panel a clear feature:  an allowed direction in parameter space that can substantially reduce $S_8$. When combining with KiDS data (light green), this is the region that is mostly preferred, favoring at $1 \sigma$ a DR temperature
\be
\xi = 0.117\substack{+0.026\\-0.016} \qquad (\text{{\it Planck} + BAO + FS + KiDS},\ f=1)\,.
\ee
Furthermore, for {\it Planck} + BAO + FS data, $\Delta \chi^2 = -5.5$ for IDM compared to $\Lambda$CDM (see Tab.\,\ref{tab:resultstension}), showing \it slight \rm preference for IDM with $f=1$ according to the AIC criterion. However, the inclusion of KiDS data leads to $\Delta \chi^2 = -11.7$, indicating a {\it strong} preference for IDM compared to $\Lambda$CDM.

From Fig.~\ref{fig:corner_lcdm_fid} we can see that the blue ({\it Planck}+BAO+FS) posteriors favor somewhat larger values of $a_{\rm dark}$ and $\xi$ compared to {\it Planck} + BAO + RSD (red), and become narrower for $\omega_{\rm cdm}$ and $H_0$.
This means that the FS analysis does provide relevant additional information within the IDM model. In addition to providing constraints for some of the cosmological parameters, FS data strongly shifts the values of IDM parameters away from the $\Lambda$CDM limit (larger IDM temperature $\xi$ and stronger interaction).

The contributions of each likelihood to the total $\chi^2$ are displayed in Tab.\,\ref{tab:chi2formodels}.  We highlight that, when including Planck and FS data, the analysis of IDM for the fiducial case improves $\chi^2$ relative to $\Lambda$CDM, but within the expected amount given the extra free parameters. This expected reduction is subtracted off for the $\Delta$AIC statistics (see Tab.\,\ref{tab:resultstension}), which does not strongly discriminate between models when leaving out weak lensing data, but slightly prefers IDM. Note that when including weak lensing (KiDS) in the likelihood for $\Lambda$CDM, the value of $\chi^2$ increases by $8.5$, most of this contribution coming from KiDS itself (and also another part coming from FS), exposing the $S_8$ tension within $\Lambda$CDM. In the case of IDM with $f=1$, $\chi^2$ only changes by 2.2 when including KiDS, with {\it Planck}  high $\ell$ being the main responsible for this shift. FS $\chi^2$ improves by about 2 when including KIDS. We can deduce that IDM accommodates KiDS and FS data way better than $\Lambda$CDM, with a small deterioration coming from the {\it Planck}  high $\ell$ fit. This indicates that an analysis including Atacama Cosmology Telescope (ACT) data and South Pole Telescope (SPT) data is an interesting direction of investigation in the future.

\begin{table}[ht]
	\small
	\begin{center}
		\begin{tabular}{ | l || c | c ||  c | c ||  c | c | }
			\hline
			\multirow{2}{*}{} & 
			\multicolumn{2}{|c||}{ $\Lambda{\rm CDM}$} &
			\multicolumn{2}{|c|}{Fiducial}  &
			\multicolumn{2}{|c|}{$f=0.1$}\\
			\cline{2-7}
			&+FS&+FS+KiDS &+FS&+FS+KiDS &+FS&+FS+KiDS \\ \hline
			
{\it Planck} High $\ell$ TT-TE-EE & 2354.3 & 2352.7 & 2348.0 & 2351.6 & 2348.7 & 2355.8 \\
{\it Planck} Low $\ell$ EE & 395.8 & 396.5 & 396.1 & 396.2 & 396.8 & 395.9 \\
{\it Planck} Low $\ell$ TT & 22.8 & 23.1 & 23.0 & 23.4 & 23.3 & 23.5 \\
{\it Planck} lensing & 9.0 & 8.9 & 9.2 & 8.7 & 9.4 & 9.5 \\
BAO    & 9.8 & 9.8 & 10.0 & 9.9 & 9.8 & 10.9 \\
FS     & 771.3 & 772.7 & 771.4 & 769.5 & 772.5 & 766.4 \\
KiDS   & - & 7.9 & - & 0.4 & - & 0.0 \\ \hline
Total  & 3563.1 & 3571.6 & 3557.7 & 3559.9 & 3560.4 & 3562.0 \\

			\hline	
			
		\end{tabular}
	\end{center}
	\caption{\small Contribution of different likelihoods described in Sec.~\ref{dataset} to the total $\chi^2$ of the best-fit for the $\Lambda$CDM, fiducial ($f=1$) and $f=0.1$ models. The values of $\chi^2$ are truncated in the first decimal, which leads to KiDS $\chi^2$ being shown as zero and the total sum being different than the sum of the parts for some models.
		\label{tab:chi2formodels}}
\end{table}

An important additional probe of scenarios suppressing the power spectrum comes from the Lyman-$\alpha$ forest. In this work we focus on the BOSS FS analysis, and leave a quantitative analysis including Lyman-$\alpha$ data to the future. However, we note that for the scenario with $n=0$ considered here, the suppression of the power spectrum is rather flat, since the DM-DR interaction rate has a similar time dependence as the Hubble rate, such that the interaction is active while a wide range of scales enter the horizon~\cite{Buen-Abad:2017gxg} (see Appendix~\ref{app:n_scaling}). This is also the reason why Lyman-$\alpha$ data are less relevant for this scenario. Indeed, in~\cite{Hooper:2022byl} it was found that {\it Planck} data are more constraining than HIRES and MIKE Lyman-$\alpha$ observations~\cite{Viel:2013fqw} for IDM.
A similar conclusion was reached in~\cite{Garny:2018byk} based on BOSS Lyman-$\alpha$ obervations~\cite{Palanque-Delabrouille:2013gaa,Chabanier:2018rga} and using a conservative analysis regarding astrophysical uncertainties. The suppression of the power spectrum is also rather gradual if only a fraction of DM interacts with DR~\cite{Chacko:2016kgg,Buen-Abad:2017gxg}. In the next section we investigate this possibility.

Before moving on, however, one more comment with respect to $H_0$ is noteworthy. The $H_0$ posteriors for $\Lambda$CDM and IDM are shown in the right panel of Fig.~\ref{fig:corner_lcdm_fid}. We see that the inclusion of FS data improves the constraints on $H_0$ for IDM when compared to {\it Planck} + BAO + RSD. The peak position does not shift towards higher $H_0$ (in the SH0ES direction \cite{Riess:2020fzl}), i.e. the $H_0$ tension is not solved by IDM, in agreement with former observations~\cite{Schoneberg:2021qvd} (though it could be mitigated when including lower bound priors on $N_{\rm eff}$~\cite{Archidiacono:2019wdp}, but with the price of making the fit worse). In our case we find 
\begin{equation}
H_0 = 68.11\substack{+0.36\\-0.37} \quad \textrm{({\it Planck} + BAO + FS, for IDM)} \,.
\end{equation}
The result for $H_0$ is similar for different values of $f$ or $n$, considered in the following sections of this work.

\subsection{Mixed cold and interacting DM} \label{sec:fchange}

The scenario in which only part of DM interacts with DR is motivated for instance by a WIMP dark sector (the PAcDM scenario of \cite{Chacko:2016kgg}) with two matter fields $\chi_1$ and $\chi_2$, in which only $\chi_2$ couples to DR \cite{Pospelov:2007mp}. $\chi_1$, in that case, remains collisionless and dominates the DM abundance. For that scenario, it is the suppression of $\chi_2$ clustering that leads to smaller $S_8$.

\begin{figure}[h]
	\centering  
	\includegraphics[width=.6\textwidth]{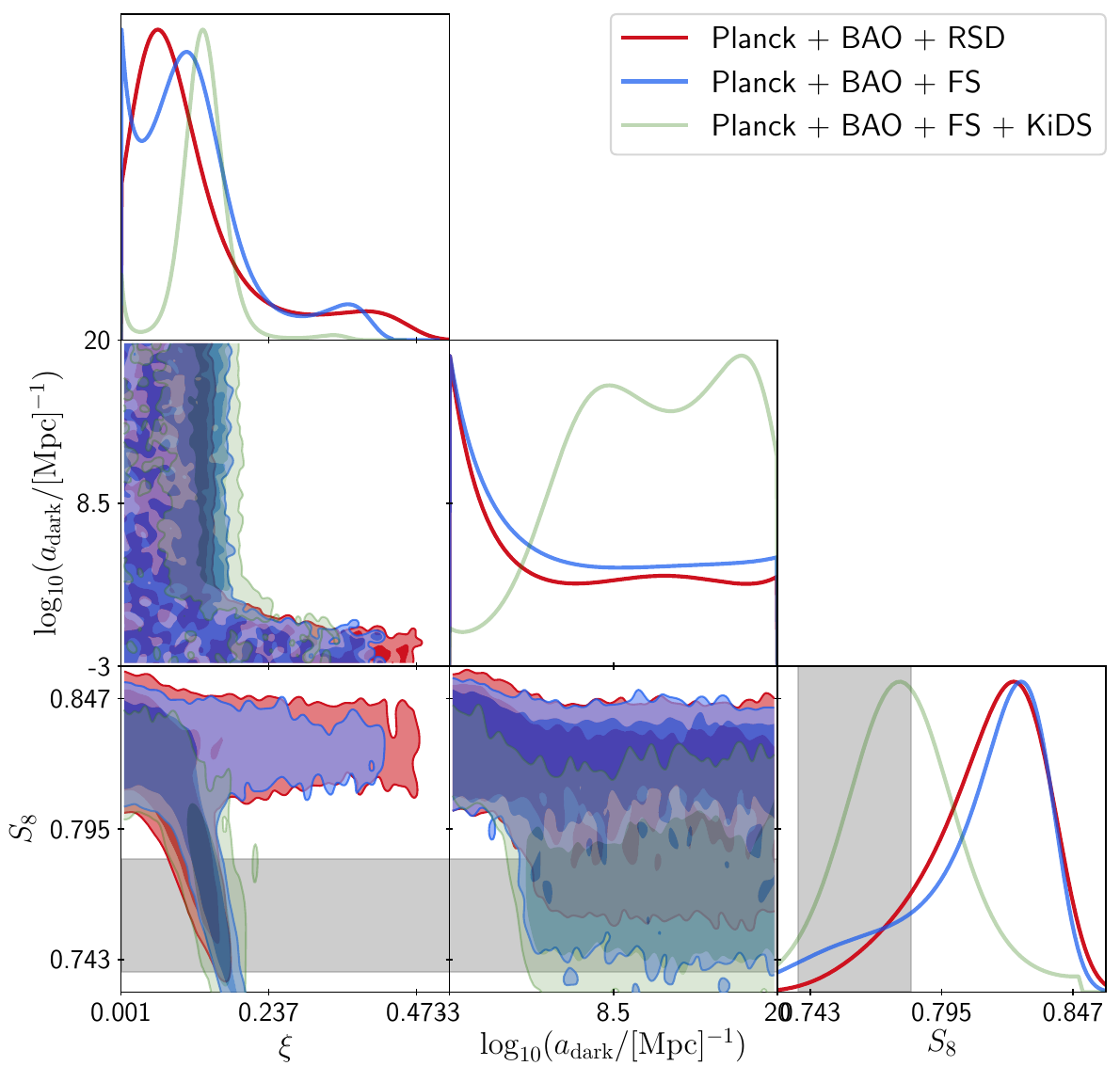}
	\caption{\label{fig:corner_f01}
		\small Same as the left panel of Fig.~\ref{fig:corner_lcdm_fid} but for IDM with interacting fraction $f=0.1$.}
\end{figure}  

We start by discussing the case in which only $10\%$ of DM interacts with DR. Notice that the ratio of IDM and CDM densities for that case is of similar magnitude as that of baryons and DM. From Tab.\,\ref{tab:resultstension}, we can see that the tension in $S_8$ is substantially alleviated to $1.3\sigma$. Also $\Delta \chi^2 = -9.6$ (with $\Delta{\rm AIC} = -5.6$) when including KiDS data, showing a preference of this model over $\Lambda$CDM. Fig.~\ref{fig:corner_f01} shows the combined posteriors for $f=0.1$. Compared to the case in which 100\% of DM interacts with DR, the 10\% case provides a good trade between reducing $S_8$ via DM-DR interaction and at the same time not inducing too much suppression and being in agreement with {\it Planck} and BAO data. Note that the $10\%$ scenario allows for strong DM-DR interaction (i.e.\ large $a_\text{dark}$) if compared to the results of Fig.~\ref{fig:corner_lcdm_fid} when including KiDS. We can connect that to the {\it dark plasma}  and {\it weakly interacting}  scenarios pointed out in Sec.~\ref{sec:ETHOS} and \cite{Buen-Abad:2017gxg}: when IDM composes $100\%$ of DM, a relatively weak DM-DR interaction strength is slightly preferred. In that case the rate is comparable to the Hubble rate while the relevant modes enter the horizon. When only $10\%$ of DM interacts, the interaction can be much stronger, leading to a tight coupling of IDM with DR, corresponding to the dark plasma scenario. Since our parameterization encompasses both limiting cases, the difference in the posteriors for $a_\text{dark}$ in Fig.~\ref{fig:corner_lcdm_fid} and in Fig.~\ref{fig:corner_f01} can be clearly associated with the weakly interacting and dark plasma limits, respectively.

Moreover, note however that when only the FS likelihood is included, there is \it no \rm preference for $10\%$ IDM ($\Delta {\rm AIC} = 1.3$). Tab.\,\ref{tab:chi2formodels} also shows the contribution of each likelihood to the total $\chi^2$ of the $10\%$ model. Note that when including KiDS, the FS likelihood strongly favors $10\%$ IDM, with some deterioration in $\chi^2$ from {\it Planck} data. The preferred value of the DR temperature when including KiDS is
\be
  \xi = 0.140\substack{+0.026\\-0.022}  \qquad (\text{{\it Planck} + BAO + FS + KiDS},\ f=0.1)\,.
\ee

\begin{figure}[h]
	\centering  
	\includegraphics[width=.45\textwidth]{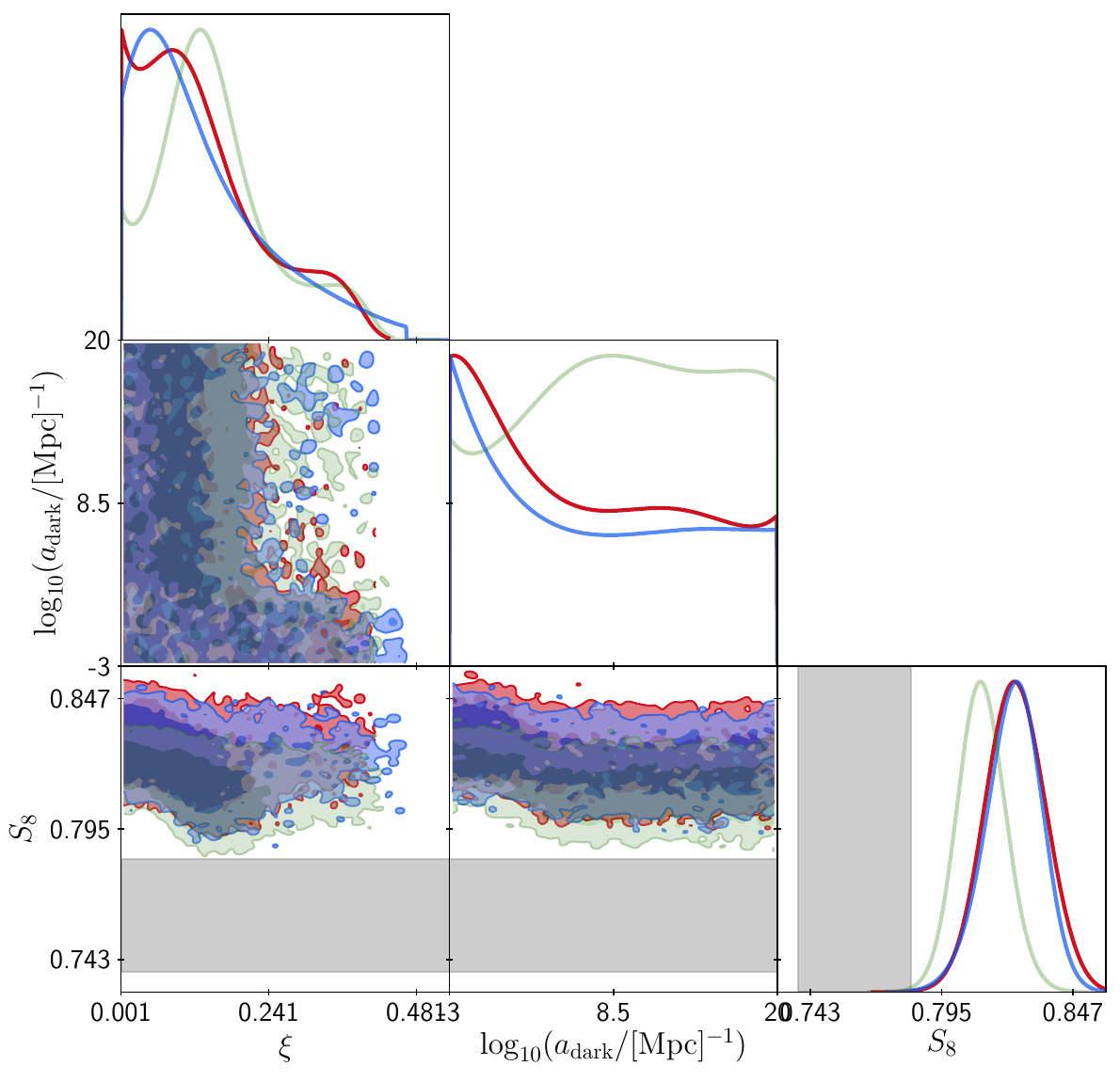}
	\includegraphics[width=.45\textwidth]{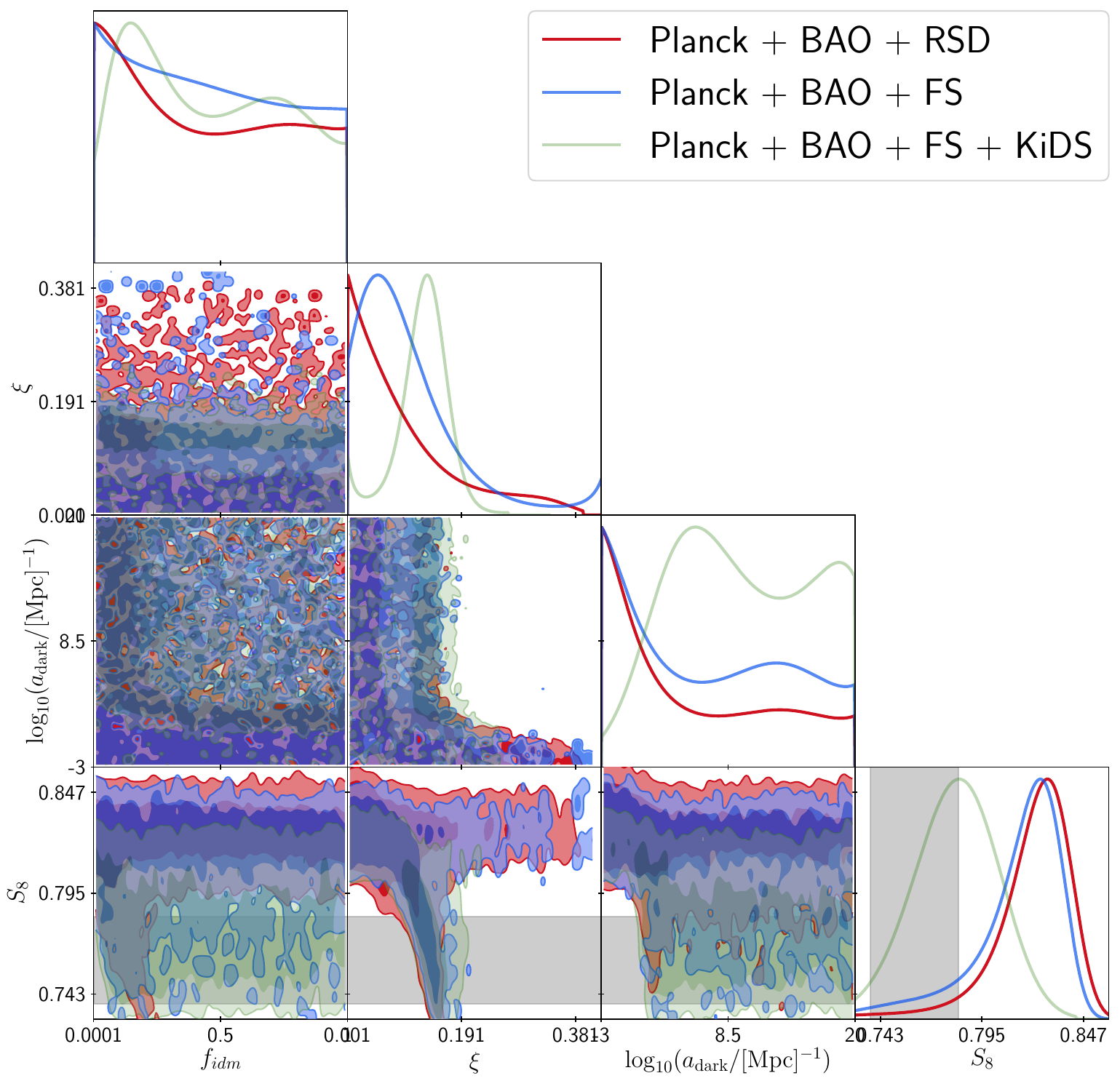}
	\caption{\label{fig:corner_fractions}
		\small Posteriors for IDM fraction $f=0.01$ (left) and for $f$ as a free parameter (right).}
\end{figure}

In the left panel of Fig.~\ref{fig:corner_fractions}, we show the MCMC results for the case in which IDM composes 1\% of DM. In this scenario, the suppression in the power spectrum is mild and not enough to shift $S_8$ towards the KiDS reference value. The tension, in that case, is only mildly alleviated to $2.3\sigma$ and $\Delta \chi^2 = -4.7$ when including KiDS, with no preference for this model.  

The right panel of Fig.~\ref{fig:corner_fractions} shows the case in which the IDM fraction $f$ is also a free parameter. We see that in that case the $S_8$ tension is alleviated to $1.9\sigma$. When including FS, $\Delta \chi^2 = -5.2$ and $\Delta {\rm AIC} = 0.8$ showing \it no \rm preference\footnote{We note that there are some fluctuations in the Monte Carlo minimization scheme for $\chi^2$, such that the case with $f$ free leads to a slightly larger value of $\chi^2$ when compared to the case $f=1$, though the second model is embedded in the first. One should therefore keep in mind that $\sim 0.5$ fluctuations are present in the total $\chi^2$. We mitigated the size of this uncertainty by running multiple independent chains as described further above. This does not change the model comparison as well as the overall conclusions that IDM substantially reduces the tension with respect to $S_8$.}. The inclusion of KiDS information induces a sharp drop of $\Delta \chi^2 = -10.2$ and $\Delta {\rm AIC} = -4.2$, favoring then IDM over CDM. Notice that the case in which $f=0$, that resembles the case of CDM plus extra DR, is {\it not} preferred over other parameter points when KiDS is included. There is even a slight preference for non-zero values for $f$, with a bump at $f\sim 10\%$, being the case already discussed above that produces good agreement with KiDS data.

\subsection{Free-streaming DR} \label{sec:freestreaming}

We now move to the free-streaming scenario for DR. In that case, we assume that the self-interaction rate of DR in (\ref{eq:gamma_dr_dr}) is negligible and the entire hierarchy of moments for the Boltzmann system is solved up to $l=17$ in CLASS. 
The free-streaming case is especially relevant for example in the case of DR being sterile neutrinos or the case in which DM is charged under an Abelian $U(1)$ symmetry. In the first scenario, the scaling of the DM-DR interaction with temperature would lead to $n=4$. For the second case, it would lead to $n=2$ \cite{Cyr-Racine:2015ihg}. We also show the $n=0$ free-streaming case.

\begin{figure}[h]
	\centering  
	\includegraphics[width=.32\textwidth]{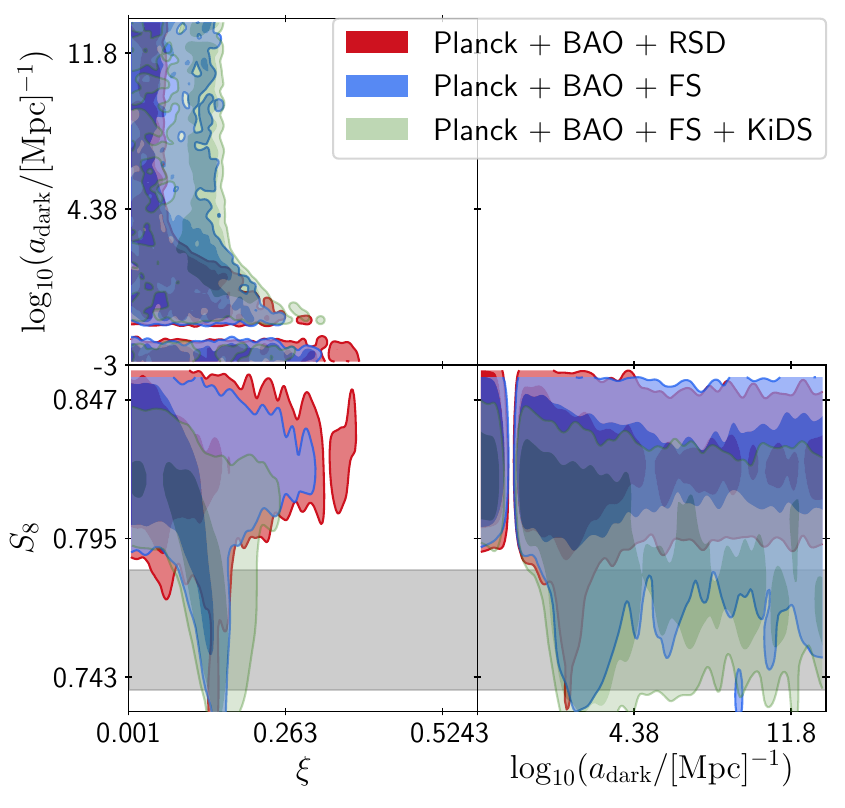}
	\includegraphics[width=.32\textwidth]{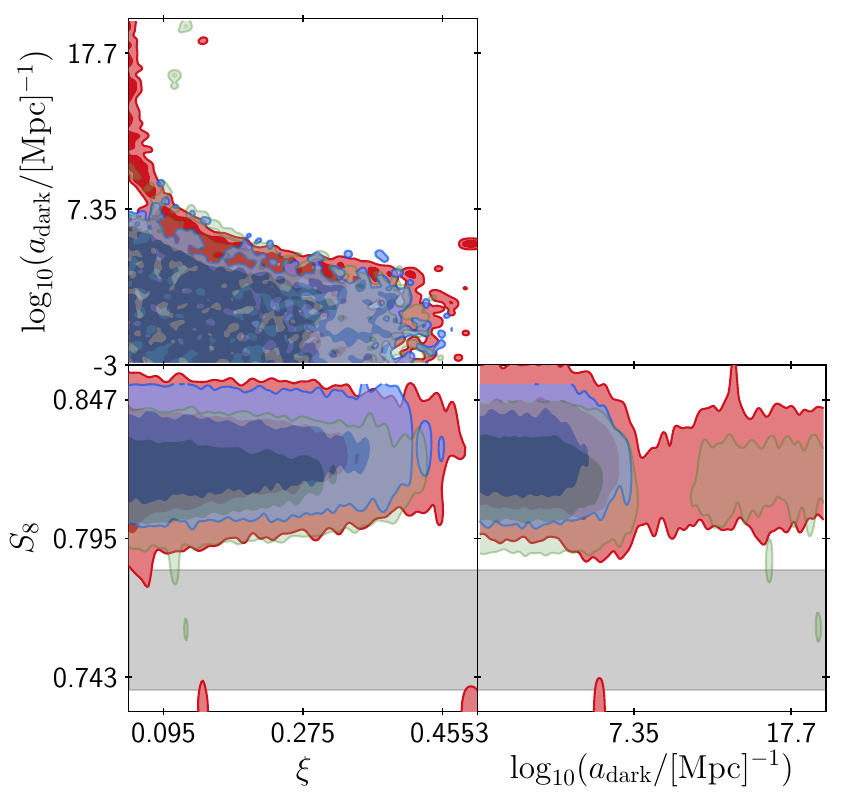}
	\includegraphics[width=.32\textwidth]{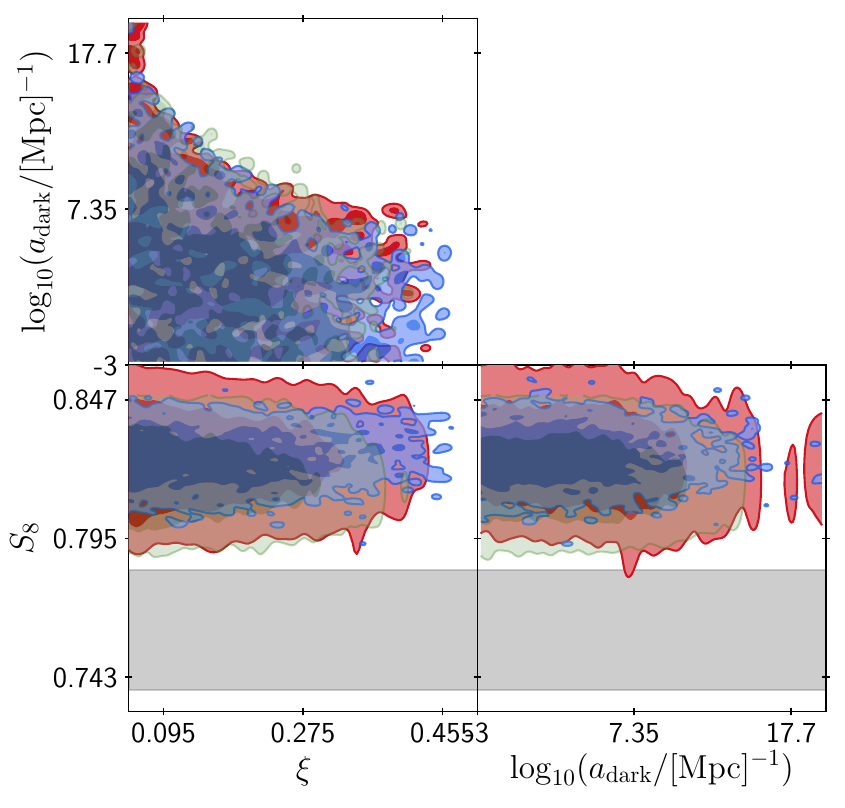}
	\caption{\label{fig:corner_n}
		\small Posteriors for the free-streaming scenario when taking $n = 0,2,4$ (left, middle, right). }
\end{figure}

The MCMC results for the free-streaming case with $n=0,2,4$ are shown in Fig.~\ref{fig:corner_n}. We see that only the  $n=0$ scenario is able to reach smaller values of $S_8$. The cases $n=2,4$ have a strong temperature dependence, leading to a sharp transition from the case in which $\Gamma/\mathcal{H} \gg 1$ to the case $\Gamma/\mathcal{H} \ll 1$ (see App.\,\ref{app:n_scaling}). As a consequence, structure formation is strongly suppressed on scales that enter the horizon before this transition. In contrast, the DM-DR interaction for $n=0$ is almost constant during radiation domination, leading to a gradual suppression of the power spectrum.

Correspondingly, we find that the parameter region compatible with low values of $S_8$ exists only for $n=0$. In that scenario, we find the smallest tension regarding KIDS data, with only $0.8\sigma$. However, as remarked before, the $Q_{\rm DMAP}$ statistic relies on finding minima, for which MCMC is not optimal, and therefore the precise number should be taken with a grain of salt. The MCMC sampling required more time to converge according to the Gelman-Rubin criterion for this scenario, and several chains were necessary. Still, the results from Tab.\,\ref{tab:resultstension} indicate a clear pattern. IDM lifts the tension between KiDS and other probes to the $\sim 1\sigma$ level, as long as $n=0$ and $f\gtrsim 0.1$.

Interestingly, for $n=2,4$, we find that adding the FS information significantly improves the upper limit on the interaction strength $a_\text{dark}$, even for rather small values of $\xi$.
This can be traced back to the scale-dependence of the power suppression for $n=2,4$, and the sensitivity of FS data to smaller scales compared to {\it Planck}. 
Furthermore, we stress an important difference between $n=0$ and $n=2,4$: in the first case, there is an allowed lower $S_8$ region that drives the preferred $\xi$ to higher values when including FS, whereas in the second case FS analysis tightens the constraints on IDM parameter space.
We also note that as opposed to $n=0$, Lyman-$\alpha$ data are expected to provide relevant additional information for $n=2,4$~\cite{Archidiacono:2019wdp}, but a combined analysis of Lyman-$\alpha$ with  {\it Planck} and FS data is beyond the scope of this work. 

\section{Dark sector with non-Abelian gauge interaction}
\label{sec:mapping}

In this section we provide an example for a microscopic DM-DR interaction model that falls into the category of models studied above.
After briefly reviewing the model, we present a computation of the DM-DR interaction rate taking Debye-screening into account, including
a non-analytic correction at next-to-leading order in the coupling expansion. We then map the previously derived constraints on the model parameter space. Furthermore, we discuss implications for small-scale structure formation as well as constraints from requiring the gauge interaction to be weakly coupled until today. 

The non-Abelian IDM-DR model~\cite{Buen-Abad:2015ova} is described by an unbroken $SU(N)$ dark gauge symmetry and a massive Dirac fermion dark matter multiplet $\chi$ in the fundamental representation,
\be
  {\cal L} = -\frac12 \text{tr}(F_{\mu\nu}F^{\mu\nu}) + \bar\chi(i\gamma^\mu D_\mu-m_\chi)\chi\,,
\ee
where $D_\mu=\partial_\mu+ig_d t^aA^a_{d,\mu}$, $F_{\mu\nu}=\frac{1}{ig_d}[D_\mu,D_\nu]$ and $A^a_{d,\mu}$ is the dark gauge field. The free parameters are, apart from $N$, the dark gauge coupling $g_d$ (or equivalently $\alpha_d=g_d^2/(4\pi)$)
and the DM mass $m_\chi$. DR is described by massless dark gauge bosons with $\eta_\text{DR}=2(N^2-1)$ degrees of freedom, while $\eta_\chi=2N$ for DM. Here we assume no interactions of DM or DR with the Standard Model that are
relevant during horizon entry of the perturbation modes probed by CMB and LSS, and during structure formation.

The model can be complemented by an additional massive field that transforms trivially under the gauge symmetry, providing a non-interacting contribution to the total dark matter density (called $\chi_1$ in Sec.\,\ref{sec:fchange}).
This option is necessary for scenarios in which only a fraction $f$ of the DM is interacting. Apart from the value of $f$ and the property of being cold and collisionless, no microscopic
information about this component is required.

The non-Abelian dark gauge boson self-interaction ensures that DR behaves as a dark fluid, described by its density contrast $\delta_\text{DR}$ and velocity divergence $\theta_\text{DR}$, while all higher multipole moments are driven to zero and taken to be vanishing.
Furthermore the temperature dependence of the DM-DR interaction rate maps onto the $n=0$ scenario (see below for details), such that the non-Abelian model coincides with the settings considered in Sec.\,\ref{sec:fiducial} and Sec.\,\ref{sec:fchange} for all or a part of DM being interacting, respectively. We also assume the gauge coupling to be sufficiently weak such that the confinement scale of the gauge interaction is much smaller than the DR temperature today, as discussed below.

\subsection{DM-DR interaction rate}
\label{sec:intrate}

\begin{figure}[h]
	\centering  
	\includegraphics[width=.95\textwidth]{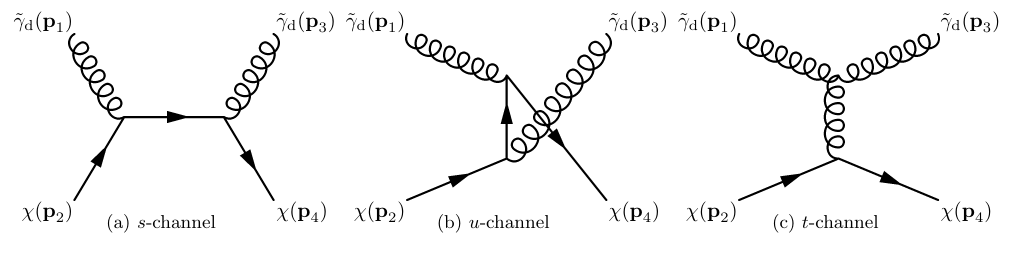}
	\caption[Feynman diagrams of DM-DR interaction]{\small Feynman diagrams for non-Abelian IDM-DR Compton scattering.}
   \label{fig:DM-DR}
\end{figure}

The interaction between DM and DR is analogous to Compton scattering of baryons and photons, with the difference of an additional $t$-channel contribution to the scattering amplitude (see Fig.\,\ref{fig:DM-DR}) involving the non-Abelian three-gluon vertex. DM-DR interactions that affect the power spectrum take place in the non-relativistic limit for the DM particle, $T_\text{DR}\ll m_\chi$.\footnote{Since typical values for $m_\chi$ are at the GeV-TeV scale while recombination occurs at the eV scale, there is a large hierarchy between modes that are affected by relativistic effects and modes probed by the CMB and LSS.} In this limit, the non-Abelian interaction leads to an important difference compared to the Abelian (Thomson) scattering process. For small scattering angles the momentum transfer in the $t$-channel goes to zero, leading to a strong enhancement of the non-Abelian contributions. Indeed, formally, the cross-section and thereby the interaction rate would diverge logarithmically when integrating over all angles.

\begin{figure}[h]
	\centering  
	\includegraphics[width=.95\textwidth]{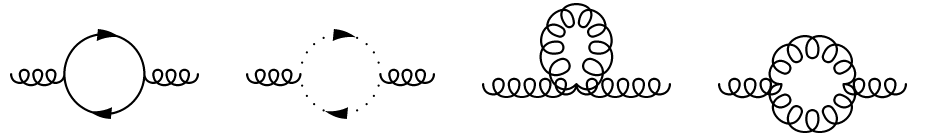}
	\caption{\small Contributions to the self-energy of the dark $SU(N)$ gauge bosons, relevant for Debye screening. Dotted lines represent ghost propagators.}
    \label{fig:self-energy}
\end{figure}

This divergence is regulated by Debye-screening in the non-Abelian plasma, described by the Debye mass~\cite{Laine:2016hma}
\be  \label{eq:debyeM}
m_D^2 =\frac{1}{3}g_d^2T_{\text{DR}}^2(C_A+\frac{1}{2}N_f)\,,
\ee
where $C_A=N$ is related to the gauge group and $N_f$ is the number of light fermionic degrees of freedom, that is zero for the relevant temperature range and within the minimal IDM-DR model considered here, but that we include below for generality. More precisely, the $t$-channel propagator needs to be replaced by the hard thermal loop (HTL) resummed propagator~\cite{Laine:2016hma}, taking the Dyson resummation of self-energy diagrams shown in Fig.\,\ref{fig:self-energy} into account. The HTL approximation captures the leading dependence in the large temperature limit, i.e. for a momentum transfer that is much smaller than $T_\text{DR}$, and taking only the dominant contribution from (ultra-)relativistic particles in the self-energy loop into account. 
The HTL approximation amounts to implementing the following replacement of the gluon propagator~\cite{Heiselberg:1994ms}
\be
t^{-1}\to (t-\Pi_{L,T})^{-1}\,,
\ee
where $L$ and $T$ refer to longitudinal and transverse projections of the gluon self-energy $\Pi_{\mu\nu}$ given e.g. in~\cite{Peshier:1998dy}, corresponding to the electric and magnetic interactions, respectively. Up to corrections that are relatively suppressed by higher powers of $T_\text{DR}/m_\chi$ only the longitudinal part contributes, and the momentum transfer is dominated by its spatial part $q\simeq (0,\vec q)$, corresponding to the Coulomb limit. Furthermore, the Debye-screening is relevant for $|\vec q|\ll T_\text{DR}$. Altogether this means that to obtain the leading dependence on $T_\text{DR}/m_\chi$, we can approximate $\Pi_{L}\to m_D^2$, while neglecting the transverse contribution.

An explicit computation taking the Debye-screening into account yields for the interaction rate (see App.\,\ref{app:details})
\footnote{It turns out that an analogous quantity exists within the quark-gluon plasma, known as heavy quark diffusion coefficient. Our result for $c_0$ matches the gluon contribution to this quantity~\cite{Moore:2004tg}. We thank Tobias Binder for pointing out this connection.}
\be\label{eq:Gamma_IDM_DR_SUN}
  \Gamma_{\rm IDM-DR} = -a\frac{\pi}{18}\frac{\alpha_d^2}{m_\chi}\eta_\text{DR}\left\{ T_\text{DR}^2 \left[ \ln\alpha_d^{-1} + c_0 + c_1g_d + {\cal O}(g_d^2)\right] + {\cal O}\left(\frac{T_\text{DR}^4}{m_\chi^2}\right)\right\}\,,
\ee
with $\alpha_d=g_d^2/(4\pi)$. The logarithmic term arises from Debye-screening and agrees with previous results~\cite{Cyr-Racine:2015ihg}. Here we provide also the constant term as well as the first-order correction with coefficients
\begin{equation}
	 c_0 = 1+\text{ln}\left(\frac{6}{2N+N_f}\right)+\text{ln}(4\pi)-24\,\text{ln}\,(A)\,, \quad \textrm{and} \quad 
	c_1 =\frac{3\sqrt{2N+N_f}}{4\pi}\sqrt{\frac{3}{2}}\,,
\end{equation}
with $A\simeq 1.28243$ being the Glaisher-Kinkelin constant. The non-analytic correction linear in $g_d$ originates from averaging over the thermal Bose-Einstein distribution of dark gauge bosons, which is strongly enhanced
for small momenta. A similar effect is well-known in the computation of various transport rates at finite temperature~\cite{Laine:2016hma}. While the logarithmic and linear terms
receive contributions only from the $t$-channel process, also the interference of $s$- and $u$-channel with the $t$-channel amplitudes contribute to the constant term $c_0$.
On the other hand, the Thomson-like contributions (being the square of the $s$- and $u$-channel as well as their interference) are suppressed by a relative factor $T_\text{DR}^2/m_\chi^2$.
We refer to App.\,\ref{app:details} for details of the computation.

\subsection{Mapping of cosmological constraints on the model parameter space}
\label{sec:parspace}

The DM-DR interaction rate~\eqref{eq:Gamma_IDM_DR_SUN} can be mapped onto the ETHOS parameterization~\eqref{eq:gamma_dm_dr}. Neglecting contributions that are suppressed in the non-relativistic limit $T_\text{DR}\ll m_\chi$
leads to $a_n=0$ for $n>0$ and $a_\text{dark}\equiv a_0$\footnote{This result differs by a factor $2$ from (37) in~\cite{Cyr-Racine:2015ihg}. However, we checked the agreement of the logarithmic contribution to $\Gamma_{\rm IDM-DR}$ with the right equation in (36) in~\cite{Cyr-Racine:2015ihg}, from which one obtains the prefactor in $a_0$ quoted here. The difference can be traced back to a missing factor of $2$ in the left equation in (36) in~\cite{Cyr-Racine:2015ihg}.}
\bea\label{eq:adarkSUN}
  a_\text{dark} &=& \frac{\pi}{12}\frac{\alpha_d^2}{m_\chi}\frac{1}{\xi_N^2}\frac{T_{\gamma,0}^2}{\Omega_\gamma h^2}\left[ \ln\alpha_d^{-1} + c_0 + c_1g_d + {\cal O}(g_d^2)\right]\nn\\
  &=& 0.91\cdot 10^9 \text{Mpc}^{-1} \left(\frac{\alpha_d}{10^{-4}}\right)^2\left(\frac{100\,\text{GeV}}{m_\chi}\right)\left(\frac{0.1}{\xi_{N}}\right)^2\left[ \ln\alpha_d^{-1} -(1.34+\ln N) +0.413\sqrt{N} g_d\right]\,,\nn\\
\eea
where $\xi_{N}=T_\text{DR}/T_\gamma$ (see below), $T_{\gamma,0}$ is the present CMB temperature, and $\Omega_\gamma h^2$ its density parameter.
In the second line we used natural units to convert the rate into the dimensions used in the previous sections. 

The DR energy density is related to its temperature according to $\rho_\text{DR}=\frac{\pi^2}{30}\eta_\text{DR}\xi_{N}^4T_\gamma^4$. 
Within the parameterization used for the MCMC analysis,
the DR temperature enters explicitly only via $\rho_\text{DR}$.
However, for the input parameter $\xi$ used in the previous sections we assumed a fiducial value
$\eta_\text{DR}^\text{fid}=2$, while for the dark $SU(N)$ model $\eta_\text{DR}=2(N^2-1)$. Nevertheless, we can map both cases by requiring that they correspond to the
same DR energy density, which implies that $\xi_{N}=\xi(N^2-1)^{-1/4}$. Here $\xi_{N}$ is related to the actual temperature of the 
$SU(N)$ dark plasma and $\xi$ is the MCMC parameter used in the sections above. Alternatively, the DR energy density can be parameterized by (see App~\ref{app:idm_extra})
\be
  \Delta N_\text{fluid} = \frac{\rho_\text{DR}}{2\frac{\pi^2}{30}\frac78 \left(\frac{4}{11}\right)^{4/3}T_\gamma^4} \simeq 
  1.11\cdot 10^{-3}\eta_\text{DR}\left(\frac{\xi_{N}}{0.15}\right)^4\,.
\ee

If the dark sector was in thermal equilibrium with the thermal bath (of the visible sector involving SM particles)
above some decoupling temperature $T_\text{dec}$, its temperature is determined by entropy conservation to be
\be
  \xi_{N}^\text{thermal} = \left(\frac{g^\text{vis}_{*s}(T)g^d_{*s}(T_\text{dec})}{g^\text{vis}_{*s}(T_\text{dec})g^d_{*s}(T_\text{DR})}\right)^{1/3}
  \simeq 0.33\left(\frac{\eta_\text{DR}+\eta_\chi}{\eta_\text{DR}}\right)^{1/3}\left(\frac{106.75}{g^\text{vis}_{*s}(T_\text{dec})}\right)^{1/3}\,,
\ee
where $s^\text{vis}=\frac{\pi^2}{45}g_{*s}^\text{vis}T^3$ and $s^d=\frac{\pi^2}{45}g_{*s}^dT_d^3$ are the entropy densities of the visible and dark
sector, respectively, and the second line assumes $T_\text{dec}>m_\chi$ and uses that $T\ll m_e, T_\text{DR}\ll m_\chi$ during the epoch when
perturbation modes relevant for CMB and LSS enter the horizon. For $T_\text{dec}<m_\chi$ the only change is that the second factor in the expression after the second equality sign is absent.
This can be compared to the required amount of DR to address the $S_8$ tension, of order $\xi_{N}\big|_{S_8}\sim 0.1(10\xi^\text{MCMC})(N^2-1)^{-1/4}$, see Sec.\,\ref{sec:results}.
Thus, if the dark sector was in thermal contact in the very early Universe, this would require additional heavy particle species within the
thermal bath to be present such that $g^\text{vis}_{*s}(T_\text{dec})\gg 106.75$. Alternatively, the perhaps
more plausible option is that the dark sector was never in thermal contact, allowing for in principle arbitrarily small values of $\xi_{N}$.

The $SU(N)$ gauge interaction becomes stronger at low energies due to the running coupling determined by the $\beta$ function
\be
  \beta(g_d) = -\frac{11}{3}C_A\frac{g_d^3}{16\pi^2} +{\cal O}(g_d^5)\,,
\ee
where $C_A=N$ and we assumed no light degrees of freedom apart from the gauge bosons for the relevant regime $T_\text{DR}<m_\chi$.
The condition that the confinement scale $\Lambda_c$ is below some maximal value $\Lambda_c^\text{max}$ can be expressed as an
upper bound on $\alpha_d$ evaluated at some reference scale $\mu_\text{ref}$ via
\be
  \alpha_d < \frac{3}{11}\frac{4\pi}{C_A}\frac{1}{\ln\frac{\mu_\text{ref}}{\Lambda_c^\text{max}}} = 0.083\frac{3}{N}\frac{1}{1+\ln[(1+z^\text{horizon\ entry})/10^6]}\,,
\ee
where we used the position of the Landau pole obtained from the one-loop beta function as a proxy of the confinement scale.
In practice, we evaluate the reference scale at the typical energy scale of the IDM-DR scattering taken to be
the DR temperature when the relevant modes $k\lesssim 1h/$Mpc for CMB and LSS start to enter the horizon, $\mu_\text{ref}=T_\text{DR}^\text{horizon\ entry}=\xi_{N}T_{\gamma,0}(1+z^\text{horizon\ entry})$ and $\Lambda_c^\text{max}=T_{\text{DR},0}=\xi_{N}T_{\gamma,0}$.

Apart from the confinement constraint on $\alpha_d$, we also include a bound on the long-range interaction strength derived from the observed ellipticity of the gravitational potential of the galaxy NGC720~\cite{Agrawal:2016quu},
\be
  \sqrt{\frac12 C_F} \alpha_d\lesssim 0.01 \frac{1}{\sqrt{f}}\left(\frac{m_\chi}{300\,\text{GeV}}\right)^{3/2}\,,
\ee
where we included a factor $C_F/2=(N^2-1)/(4N)$ relative to the Abelian case obtained from the color factor for $\chi\chi\to\chi\chi$ scattering and averaging over the dark $SU(N)$ degrees of freedom of one of the incoming $\chi$, while summing over all others, as appropriate for the scaling of the scattering rate of a given $\chi$ particle. In addition, we estimate the change of the bound if only a fraction $f$ of DM is interacting by rescaling it according to the decrease of the scattering rate.
The magnitude of the self-scattering cross-section can be estimated as~\cite{Agrawal:2016quu}
\be
  \frac{\sigma}{m_\chi}=\frac{C_F}{2}\frac{8\pi\alpha_d^2}{m_\chi^3 v^4} \sim 1 \frac{\text{cm}^2}{g} \frac{C_F}{2} \left(\frac{\alpha_d}{2.5\cdot 10^{-5}}\right)^2 \left(\frac{100\,\text{GeV}}{m_\chi}\right)^3 \left(\frac{30\text{km}/\text{s}}{v}\right)^4\,,
\ee
where $v$ is the relative velocity. Self-interaction cross-sections of order cm$^2$/g are considered in the context of small-scale puzzles, specifically the core-cusp problem~\cite{Tulin:2017ara}.

\begin{figure}[h]
	\centering  
	\includegraphics[width=.48\textwidth]{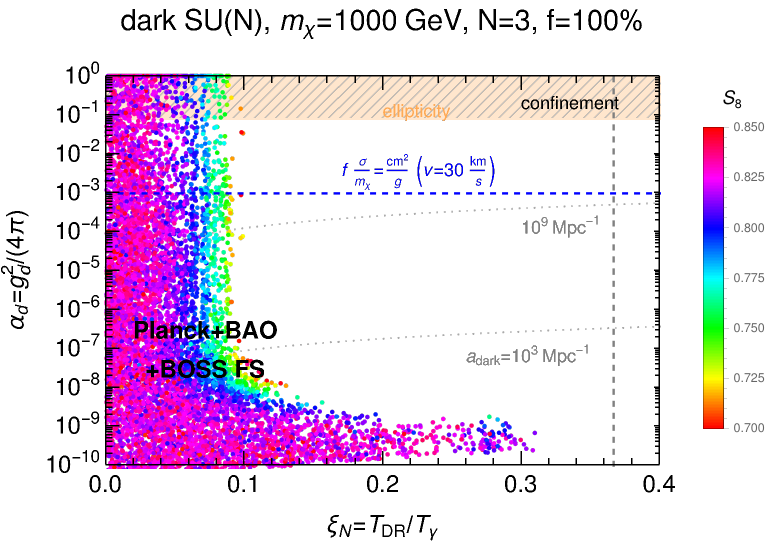}
	\includegraphics[width=.48\textwidth]{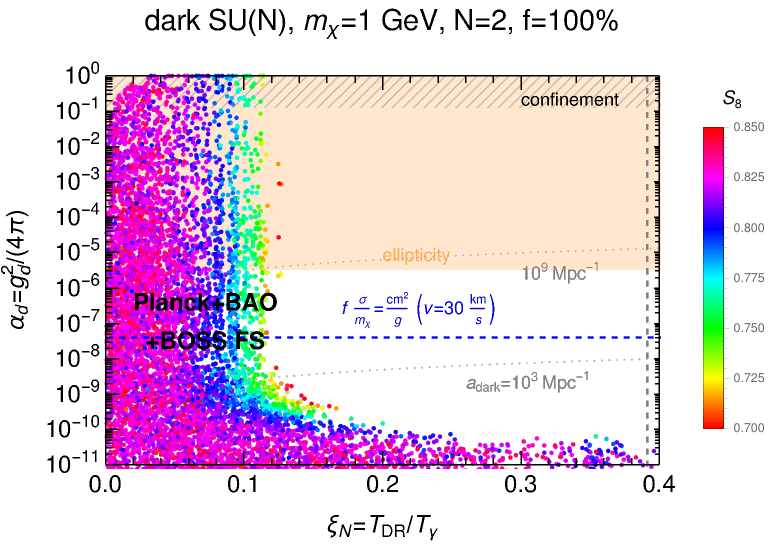}
         \\[2ex]
	\includegraphics[width=.48\textwidth]{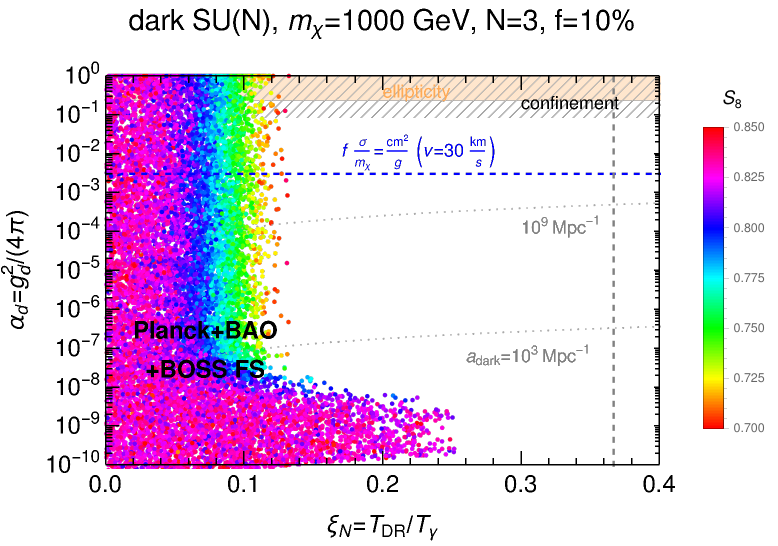}
	\includegraphics[width=.48\textwidth]{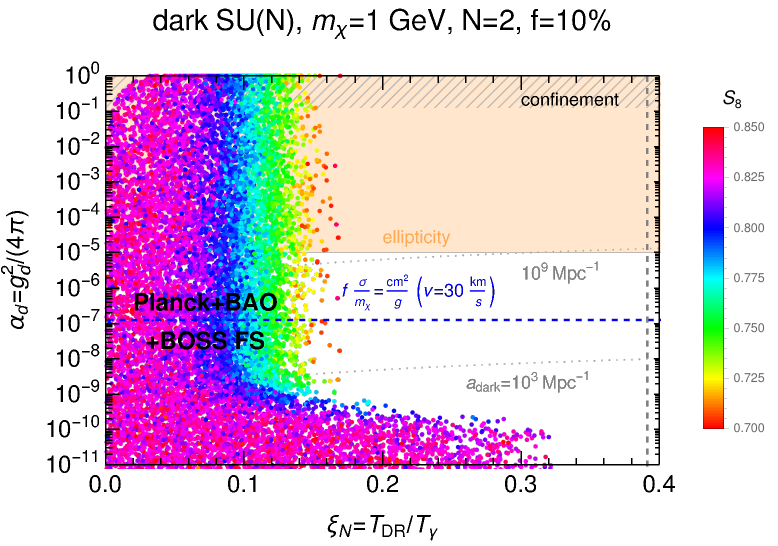}
	\caption{\label{fig:alphavsxi}	\small 
	Dark fine-structure constant $\alpha_d$ versus dark sector temperature ratio $\xi_{N}=T_\text{DR}/T_\gamma$. Coloured points are IDM-DR models allowed by {\it Planck}+BAO+BOSS FS data at $95\%$C.L., colored according to their value of $S_8$. The hatched region is excluded from the condition that the gauge force is unconfined until today, and the orange shaded region from the ellipticity of NGC720. The blue dashed line corresponds to a self-interaction cross-section of $f\times \sigma/m_\chi=1\,\text{cm}^2/\text{g}$. The vertical grey dashed line indicates the DR temperature if the dark sector was in thermal contact with the Standard Model in the very early Universe. Dotted contour lines correspond to $\log_{10}(a_{\text{dark}} [\text{Mpc}^{-1}])=3,\,9$. We show different values of $m_\chi=1000,1$\,GeV, $N=3,2$ and $f=1,0.1$ for illustration.
	}
\end{figure}

We use the result~\eqref{eq:adarkSUN} for the IDM-DR rate to map the {\it Planck}+BAO+FS constraints obtained in Sec.\,\ref{sec:fiducial} and Sec.\,\ref{sec:fchange} onto the parameter space of the
dark fine-structure constant $\alpha_d=g_d^2/(4\pi)$ versus the DR temperature $\xi_{N}$. The results are shown in Fig.\,\ref{fig:alphavsxi}, assuming $m_\chi=1$TeV and $N=3$ (left panels)
or $m_\chi=1$GeV and $N=2$ (right panels) for illustration. In addition, we show results for $f=100\%$ (upper panels) as well as $f=10\%$ (lower panels). Each point represents a set of model parameters that is
consistent with {\it Planck} + BAO + FS at $95 \%$ C.L. The points are colored according to the value of $S_8$. We see that for DR temperatures above $\xi_{N}\gtrsim 0.1-0.15$ there are stringent constraints on
the allowed gauge coupling, of order $\alpha_d\lesssim 10^{-8}-10^{-10}$. For larger couplings, the DM-DR interaction would wash out structures on scales that are very well constrained by CMB and LSS data.
However, the constraints on $\alpha_d$ greatly relax for smaller DR temperatures. In that case, the energy density of the DR bath is too small to have a sizeable impact on the power spectrum.

The region in parameter space that is favoured when comparing to the KiDS results \cite{KiDS:2020suj} $S_8 = 0.759\substack{+0.024\\-0.021}$ occurs for DR temperatures of order $\xi_{N}\sim 0.1$, with slightly lower
values for $N=3$ compared to $N=2$ due to the larger number of gauge bosons in the former case. Furthermore, for $f=100\%$ there is a tendency to prefer smaller values of $\alpha_d$ for points with $S_8$ in the KiDS range, although large values
are also possible. For $f=10\%$, the interaction strength is only bounded from below by $\alpha_d\gtrsim 10^{-8} (10^{-9})$ for $m_\chi=1000 (1)$\,GeV when requiring $S_8$ compatible with KiDS, in accordance with the limiting case of
a tightly coupled dark plasma.

In addition, the mapping allows us to compare complementary constraints from confinement and the impact of long-range interactions on elliptical galaxies, as discussed above. The corresponding exclusion regions are shown as hatched and orange shaded areas in Fig.\,\ref{fig:alphavsxi}, respectively. We see that ellipticity bounds are more constraining for light dark matter masses, and confinement for heavy masses in the TeV regime. Nevertheless, these constraints are compatible with points in parameter space that are allowed by cosmological {\it Planck}+BAO+FS data and have $S_8$ values favored by KiDS. In addition, the allowed parameter space is compatible with a self-interaction cross-section of order 1 cm$^2$/g, being relevant in the context of small-scale structure puzzles.

As anticipated above, the DR temperature favored by KiDS is significantly smaller than the one that would be expected if the dark and visible sectors would have been in thermal equilibrium in the very early Universe. This points to a dark sector that was never in thermal equilibrium with the Standard Model. A possible production mechanism of dark sector particles, in that case, is via freeze-in~\cite{Hall:2009bx}. In a minimal setup, freeze-in can occur via gravitational interactions only~\cite{Garny:2015sjg}, naturally leading to a dark sector temperature that is significantly smaller than the SM, depending on the reheating temperature. The further evolution of the dark sector depends on whether the dark gauge coupling is strong enough to establish thermal equilibrium~\cite{Garny:2018grs}. If that is the case, the annihilation $\chi\chi\to\ A_dA_d$ can lead to a further depletion of the density of $\chi$ particles, and its freeze-out within the dark sector determines the DM abundance~\cite{Agrawal:2016quu}. A similar evolution occurs if the dark and visible sectors interact via higher-dimensional operators suppressed by some large energy scale. This interaction would leave the phenomenology of structure formation discussed here unchanged, but could set the initial DR temperature and DM abundance via UV freeze-in~\cite{Elahi:2014fsa} and subsequent dark sector thermalization and freeze-out~\cite{Forestell:2018dnu}. In all these scenarios the DR temperature and DM abundance are related to the reheating temperature, which is in turn related to the scale of inflation and thereby to the tensor-to-scalar ratio $r$. In particular, gravitational production typically requires a high reheating/inflation scale, leading to a lower bound on $r$, potentially testable with upcoming CMB B-mode polarization measurements~\cite{Errard:2015cxa}. We leave a further exploration of DM and DR production mechanisms to future work. 

\section{Conclusion}
\label{sec:conslusion}

In this work we have derived constraints on dark matter interacting with dark radiation based on the full shape (FS) information of the redshift-space galaxy clustering data from BOSS-DR12, combined with BAO and {\it Planck} legacy temperature, polarization, and lensing data.
In addition, we have quantified the extent to which the $S_8$ tension can be relaxed due to DM-DR interactions, by taking the value of $S_8$ measured by KiDS into account.
We have considered a range of scenarios using the ETHOS parameterization, that differ in the temperature dependence of the interactions rate ($n=0,2,4$), the assumptions on DR self-interactions (dark fluid vs.\ free-streaming), and the fraction $f$ of DM that interacts with DR ($f=100\%,\, 10\%,\, 1\%$ or free). 

We find that, for IDM-DR models, FS data add some information when compared to the case of using only measurements of the clustering amplitude and growth rate $f\sigma_8$ extracted from redshift space distortions. Furthermore, we find that {\it Planck} and FS data are compatible with low values of $S_8$ preferred by KiDS for models with approximately constant $\Gamma/\mathcal{H}$ ($n=0$). We use several statistical indicators to assess the ability to reduce the $S_8$ tension. According to the $Q_\text{DMAP}$ statistic the tension between {\it Planck}+FS and KiDS can be reduced from $2.9\sigma$ for $\Lambda$CDM to a level of $\sim1\sigma$ for an IDM-DR model with $n=0$, and fraction $f\gtrsim 10\%$ of interacting DM, and a DR temperature $\xi\sim 0.1$ about an order of magnitude below the CMB temperature. 

The class of scenarios favored by relaxing the $S_8$ tension can be mapped on a microscopic model with a dark sector that interacts via an unbroken, weakly coupled $SU(N)$ gauge symmetry that was never in thermal equilibrium with the Standard Model.
We compute the relevant interaction rate taking higher order corrections from Debye screening into account, and map the constraints on the parameter space of the $SU(N)$ model. We find that a solution of the $S_8$ tension requires a dark fine-structure constant $\alpha_d\gtrsim 10^{-8} (10^{-9})$ for DM mass $m_\chi=1000 (1)$\,GeV. This lower bound is compatible with upper bounds from the asphericity of elliptical galaxies, and allows for self-interaction cross-sections relevant for small-scale structure puzzles.
It would be interesting to work out a production mechanism and thermal history of a dark sector with properties hinted at by the $S_8$ measurements from {\it Planck}, BOSS and KiDS, for example along the lines of gravitational or UV-dominated freeze-in, which is left for future work. In addition, it will be interesting to consider complementary probes of large-scale structure, such as the cross-correlation of weak lensing and galaxy clustering or galaxy cluster number counts.

\section*{Acknowledgment}

The authors thank Tobias Binder, Petter Taule, Sebastian Bocquet and Joe Mohr for helpful discussions. 
We acknowledge support by the Excellence Cluster ORIGINS, which is funded by the Deutsche Forschungsgemeinschaft (DFG, German Research
Foundation) under Germany’s Excellence Strategy - EXC-2094 - 390783311.

\appendix

\section{Additional information about IDM} \label{app:idm_extra}

We dedicate this appendix to reviewing some properties of the ETHOS framework, highlighting some well-known features that are instructive for understanding the difference between the various ETHOS scenarios.

\subsection{The scaling of the IDM-DR coupling} \label{app:n_scaling}

In order to understand when the IDM-DR coupling is relevant, we can start from (\ref{eq:gamma_dm_dr}) and write (assuming that only a single term in the $n$ sum is relevant)
\begin{eqnarray} 
\Gamma_{\rm IDM-DR} &=& - \frac{4}{3}(\Omega_{\rm DR}h^2)\,x_{\rm DM}(z) a_n  \frac{(1+z)^{n+1}}{(1+z_{\rm D})^n} \nonumber \\
 &=& 
 - \frac{4}{9} \frac{h^2}{H_0^2M_{\rm Pl}^2} \,\rho_{\rm DR}^0 \,x_{\rm DM}(z) a_n  \frac{(1+z)^{n+1}}{(1+z_{\rm D})^n} \nonumber \\
 &=& 
 - \frac{4}{9} \frac{h^2}{H_0^2M_{\rm Pl}^2} \,\rho_{\gamma}^0 \xi^4 \,x_{\rm DM}(z) a_n  \frac{(1+z)^{n+1}}{(1+z_{\rm D})^n} \,,
\end{eqnarray}
where the superscript $0$ denotes quantities today and $M_{\rm Pl}$ is the Planck mass. We find then
\begin{eqnarray} 
\frac{\Gamma_{\rm IDM-DR}}{\mathcal{H}} &=& 
 \frac{ \frac{4}{9} \frac{h^2}{H_0^2} M_{\rm Pl}^{-2}\,\rho_{\gamma}^0 \xi^4 \,x_{\rm DM}(z) a_n \frac{(1+z)^{n+2}}{(1+z_{\rm D})^n} }{H_0 \left[\Omega_m (1+z)^{3} + \Omega_\gamma(1 + \xi^4 ) (1+z)^{4} + \Omega_\Lambda \right]^{1/2} } \propto (1+z)^n \,.
\end{eqnarray}
Here we assumed for simplicity that DR is bosonic ($\zeta = 1$) and has two polarizations ($\eta_{\rm DR} = 2$). Substituting typical values for the constants, we obtain
\begin{eqnarray} 
 \frac{\Gamma_{\rm IDM-DR}}{\mathcal{H}} &\simeq& 
 0.0152 \left(\frac{a_n}{1000 \, \textrm{Mpc}^{-1}}\right) \left(\frac{\xi}{0.1}\right)^{4}  P(z,\xi) \,,
\end{eqnarray}
with 
\begin{equation}
P(z,\xi) \equiv \frac{ x_{\rm DM}(z)  \frac{(1+z)^{n+2}}{(1+z_{\rm D})^n} }{ \left[\Omega_m (1+z)^{3} + \Omega_\gamma(1 + \xi^4 ) (1+z)^{4} + \Omega_\Lambda \right]^{1/2} } \,,
\end{equation}
where the dependence of $P$ on $\xi$ is subdominant since typically $\xi^4 \ll 1$.

\begin{figure}[h]
\centering  
  \includegraphics[width=.6\textwidth]{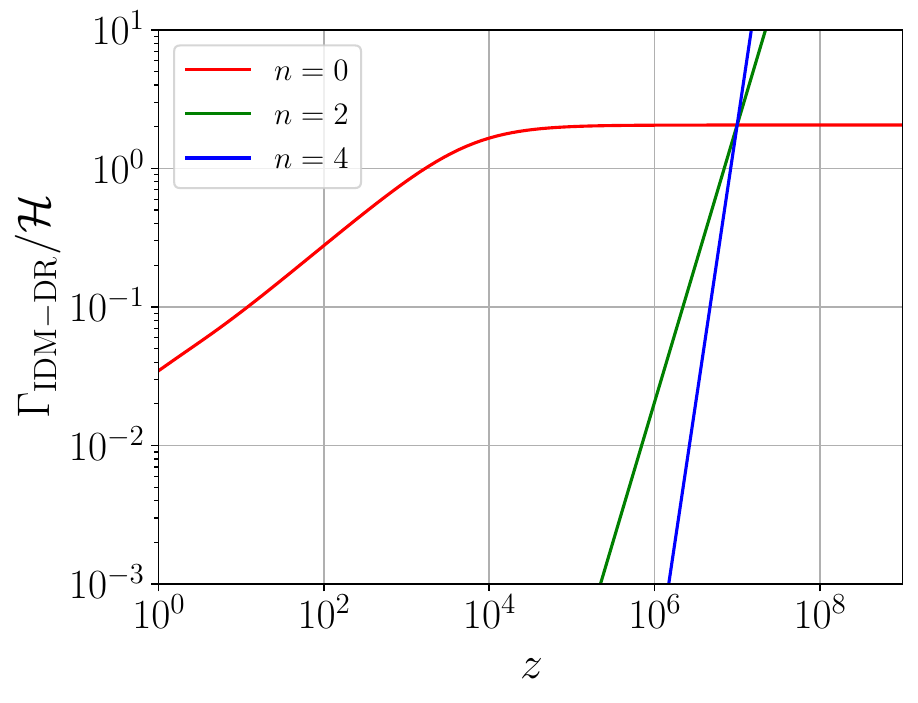}
\caption{\label{fig:gamma_h}
\small Ratio between $\Gamma_{\rm IDM-DR}$ and $\mathcal{H}$ as a function of redshift $z$ for different $n$.}
\end{figure}

We display $\Gamma_{\rm IDM-DR}/\mathcal{H}$ in Fig.~\ref{fig:gamma_h} for $n=0,\, 2, \, 4$, $\xi=0.1$ and $a_n=1000$Mpc$^{-1}$. We can see that the choices $n = 2,\,4$ lead to a sharp decrease in the interaction rate with time (i.e.\ in the direction of lower redshift). In practice, this means that only modes that enter the horizon before the ratio falls below unity are affected by the interaction, leading to a sharp $k$-dependent suppression of the power spectrum on small scales. This is the reason why these scenarios cannot account for the $S_8$ tension, which requires a more shallow suppression of the power spectrum. In addition, it implies that model parameters where the suppression occurs on scales that are smaller than those probed by CMB and LSS data cannot be tested in that way, except for the usual impact of the extra DR density.
In contrast, the $n=0$ scenario exhibits an interaction rate that evolves in time in the same way as the Hubble rate, during radiation domination. That implies that the interaction is relevant while a wide range of scales enter the horizon during the radiation era, and correspondingly leads to a rather flat suppression of the matter power spectrum.

\subsection{Mapping $\xi$ onto $\Delta N$} \label{app:xi_delta_mapping}

In this section we review the relation between the distinct ways to parametrize the energy density of dark radiation. The total energy density for radiation
\begin{equation}
\rho_r = \rho_\gamma + \rho_\nu = \left[ 1 + \frac{7}{8}\left( \frac{4}{11}\right)^{4/3}N_{\rm eff} \right] \rho_\gamma \,,
\end{equation}
can be generalized to include extra species as
\begin{equation} \label{eq:rho_rad_dr}
\rho_r = \rho_\gamma + \rho_\nu + \rho_{\rm DR} = \left[ 1 + \frac{7}{8}\left( \frac{4}{11}\right)^{4/3}N_{\rm eff} + \frac{\eta_{\rm DR}}{2} \zeta \xi^4 \right] \rho_\gamma \,,
\end{equation}
where $N_{\rm eff}$ is the effective number of neutrino species and we used that
\begin{equation}
\rho_{\rm DR} = \eta_{\rm DR} \zeta \pi^2 \frac{T^4_{\rm DR}}{30} = \frac{\eta_{\rm DR}}{2} \zeta \xi^4 \rho_{\gamma}\,.
\end{equation}
Eq.~(\ref{eq:rho_rad_dr}) can also be written as 
\begin{equation}
\rho_r = \left[ 1 + \frac{7}{8}\left( \frac{4}{11}\right)^{4/3}\left(N_{\rm eff} + \Delta N_{\rm eff}\right) \right] \rho_\gamma \,,
\end{equation}
and for the case of a dark fluid the naming $\Delta N_{\rm fluid} = \Delta N_{\rm eff}$ is commonly used. Adopting this nomenclature, we can identify
\begin{equation}
 \Delta N_{\rm fluid} = \frac{\rho_{\rm DR}}{\rho_{\rm 1\nu}} = \frac{\eta_{\rm DR}}{2} \zeta \xi^4  \frac{8}{7}\left( \frac{11}{4}\right)^{4/3} \simeq 4.4030 \, \xi^4\, \,,
\end{equation}
as the ratio between the DR energy density and the energy density of one neutrino family. In the last line we inserted $\zeta = 1$ and $\eta_{\rm DR} = 2$ as example.

\section{Computational details of DM-DR interaction}
\label{app:details}

This appendix is dedicated to presenting some of the computational details of the DM-DR interaction rate within the $SU(N)$ model. 

As already stated, the non-Abelian DM-DR interaction ($\Tilde{\gamma}_\text{d}(\textbf{p}_1)+\chi(\textbf{p}_2)\rightarrow\Tilde{\gamma}_\text{d}\textbf{p}_3)+\chi(\textbf{p}_4)$) has three distinct contributions shown by Feynman diagrams in Fig.\,\ref{fig:DM-DR}, where $P_i$ is the $i^{th}$ four-momentum and $\textbf{p}_i$ is the $i^{th}$ three-momentum. The first ingredient to compute is the squared amplitude, averaged over initial states and summed over final states. After evaluating the color numbers and the traces of gamma matrices using FORM \cite{Vermaseren:2000nd}, the result of each contribution is
\begin{eqnarray}
		\overline{|\mathcal{M}_s|^2}=&\frac{(N^2-1)}{N^2}\frac{g_d^4}{2}\,2\bigg(2m_\chi^4+m_\chi^2(s-m_\chi^2)-\frac{1}{2}(s-m_\chi^2)(u-m_\chi^2)\bigg)\frac{1}{(s-m_\chi^2)^2}\,, \\
		\overline{|\mathcal{M}_u|^2}=&\frac{(N^2-1)}{N^2}\frac{g_d^4}{2}\,2\bigg(2m_\chi^4+m_\chi^2(u-m_\chi^2)-\frac{1}{2}(s-m_\chi^2)(u-m_\chi^2)\bigg)\frac{1}{(u-m_\chi^2)^2}\,,\\
		\overline{|\mathcal{M}_t|^2}=&2g_d^4(s-m_{\chi}^2)(m_{\chi}^2-u)\frac{1}{t^2}\\
		2Re\,\overline{\mathcal{M}_s\mathcal{M}_u^*}=&\frac{-1}{N^2}\frac{g_d^4}{2}2\bigg(4m_{\chi}^4+m_{\chi}^2(s-m_\chi^2)+m_\chi^2(u-m_\chi^2) \bigg)\frac{1}{(s-m_\chi^2)(u-m_\chi^2)}\,,\\
		2Re\,\overline{\mathcal{M}_t\mathcal{M}_s^*}=&-g_d^4\bigg(-(s-m_\chi^2)^2-(s-m_\chi^2)t+m_{\chi}^2(u-m_\chi^2) \bigg) \frac{1}{(s-m_\chi^2)t}\,,\\
		2Re\,\overline{\mathcal{M}_t\mathcal{M}_u^*}=&-g_d^4\bigg(-(u-m_\chi^2)^2-(u-m_\chi^2)t+m_{\chi}^2(s-m_\chi^2) \bigg) \frac{1}{(u-m_\chi^2)t}\,,
\end{eqnarray}
with $s$, $u$ and $t$ being the Mandelstam variables. Adding everything together and evaluating at $t=2p_1^2(\Tilde{\mu}-1)$ and $s=m_{\chi}^2+2p_1m_{\chi}$ (with $\tilde\mu=\Hat{\textbf{p}}_1\cdot  \Hat{\textbf{p}}_3$, $\Hat{\textbf{p}}_i$ being unit vectors, and $p_1=|\textbf{p}_1|$), we get the total squared amplitude
\begin{equation}\label{eq:totalM}
\begin{split}
	     \left(\frac{1}{\eta_{\chi}\eta_{\text{DR}}}\sum |\mathcal{M}|^2 \right)\biggl|_{\subalign{t&=2p_1^2(\Tilde{\mu}-1)\\s&=m_{\chi}^2+2p_1m_{\chi}}}
     		\simeq{}&\frac{2g_d^4\, 4p_1^2m_{\chi}^2}{(2p_1^2(\Tilde{\mu}-1)-m_{\text{D}}^2)^2}-\frac{1}{2N^2}g_d^4(1+\Tilde{\mu}^2)\\
	+&g_d^4\,\bigg(2+\frac{m_{\chi}^2}{p_1^2}+\frac{4m_{\chi}^2}{2p_1^2(\Tilde{\mu}-1)-m_{\text{D}}^2}\bigg)\,,  \hspace{0.5cm} \text{for} \hspace{0.5cm}p_1\ll m_{\chi}\,. 
\end{split}
\end{equation}
 Note that we used here the corrected propagator of the gluon, meaning $t\rightarrow t-m_{D}^2$ in the denominator (see the discussion in Sec.\,\ref{sec:intrate} for a definition of the Debye mass $m_D$). The first term in \eqref{eq:totalM} comes from the $t-$channel alone, the second term is equivalent to the Thomson scattering in the Abelian case, and the last term is the sum of what is left from the $s/u-$channels and their interference with the $t-$channel. The approximation sign comes from neglecting terms proportional to $m_\chi p_1^3 (\tilde\mu-1)$ in the numerator of the $t-$channel contribution, being the first expression on the right-hand side. They are suppressed in the non-relativistic limit $p_1\lesssim T_{\rm DR}\ll m_\chi$.

\subsection{Interaction rate (DM drag opacity)} \label{App:drag_opacity}

The interaction rate in the ETHOS formalism is given by (for more details see Appendix A of \cite{Cyr-Racine:2015ihg})
\be 
    \Gamma_{\rm IDM-DR}\equiv{}a\frac{1}{16\pi m_{\chi}^3}\frac{\eta_{\text{DR}}}{3}\int\frac{p_1^2\text{d}p_1}{2\pi^2}p_1^2\frac{\partial\,f_{\text{DR}}^{(0)}(p_1)}{\partial\,p_1}\left(A_0(p_1)-A_1(p_1)\right)={}\frac{4\rho_{\text{DR}}}{3n_{\chi}^{(0)}m_{\chi}}\Gamma_{\text{DR}-\text{IDM}}\,,
\ee
where $f_{\text{DR}}^{(0)}(p_1)=(e^{p_1/T_\text{DR}}-1)^{-1}$ is the unperturbed DR distribution function, $n_{\chi}^{(0)}$ is the spatially homogeneous DM number density, and $\rho_{\rm DR}=\eta_{\rm DR} \zeta \pi^2 T^4_{\rm DR}/30$ is the homogeneous part of the DR energy density (with $\zeta=1$ for bosonic DR and $\zeta=7/8$ for fermionic DR). In order to compute $\Gamma_{\rm IDM-DR}$, one needs first to compute the coefficients $A_0(p_1)$ and $A_1(p_1)$ using the general formula for the projection of the spin-summed squared matrix element onto the $l^{\text{th}}$ Legendre polynomial $P_l(x)$
\be
 A_l(p)={}\frac{1}{2} \int_{-1}^1\text{d}\Tilde{\mu}\,P_l\,(\Tilde{\mu})\left(\frac{1}{\eta_{\chi}\eta_{\text{DR}}}\sum |\mathcal{M}|^2 \right)\biggl|_{\subalign{&t=2p_1^2(\Tilde{\mu}-1)\\ &s=m_{\chi}^2+2p_1m_{\chi}}}\,,
\ee
with $P_0(\Tilde{\mu})={}1$ and $P_1(\Tilde{\mu})=\Tilde{\mu}$.

Following the result we got for the full squared amplitude in~\eqref{eq:totalM}, we split the computation of the interaction rate into two parts. We start by the $t-$channel term, indicated with a superscript $t$
\begin{subequations}
    \begin{align}
        A_0^{t}(p_1)={}&g_d^4 m_{\chi}^2\frac{8p_1^2}{m_{\text{D}}^2(m_{\text{D}}^2+4p_1^2)}\,,\\
        A_1^{t}(p_1)={}&g_d^4m_{\chi}^2\left[\frac{8p_1^2}{m_{\text{D}}^2(m_{\text{D}}^2+4p_1^2)}+\frac{4}{m_{\text{D}}^2+4p_1^2}+\frac{1}{p_1^2} \left(\text{ln}\Big(\frac{m_{\text{D}}^2}{T^2_{\text{DR}}}\Big)-\text{ln}\Big(\frac{m_{\text{D}}^2+4p_1^2}{T^2_{\text{DR}}}\Big) \right)\right]\,,
    \end{align}
\end{subequations}
leading to the following contribution to the interaction rate from the square of the $t-$channel diagram,
\begin{eqnarray}\label{eq:Gamma1unintegrated}
\Gamma_{\rm IDM-DR}^{t}=C m_{\chi}^2\int&& \frac{p_1^2\text{d}p_1}{2\pi^2}p_1^2\frac{\partial f_{\text{DR}}^{(0)}(p_1)}{\partial p_1} \times   \\ && \bigg[\frac{1}{p_1^2}\bigg(-\text{ln}\Big(\frac{m_{\text{D}}^2}{T^2_{\text{DR}}}\Big)-\text{ln}\Big(\frac{T^2_{\text{DR}}}{4p_1^2}\Big)+\text{ln}\Big(\frac{m_{\text{D}}^2}{4p_1^2}+1\Big) \bigg)
-\frac{4}{m_{\text{D}}^2+4p_1^2}\bigg]\,.	\nonumber
\end{eqnarray}
Here we defined $C=a\frac{1}{16\pi m_{\chi}^3}\frac{\eta_{\text{DR}}}{3}g_d^4$ for convenience. This result can be further split into four parts. The contribution from the first two terms in the integrand of~\eqref{eq:Gamma1unintegrated} can be computed analytically. We therefore put them together and perform the integration by parts to obtain
\be\begin{split}\label{eq:i}
i + ii ={}&C m_{\chi}^2\int \frac{p_1^2\text{d}p_1}{2\pi^2}\frac{\partial f_{\text{DR}}^{(0)}(p_1)}{\partial p_1}\left(-\text{ln}\Big(\frac{m_{\text{D}}^2}{T^2_{\text{DR}}}\Big)-\text{ln}\Big(\frac{T^2_{\text{DR}}}{4p_1^2}\Big)\right)\\
={}&-C m_{\chi}^2\frac{1}{6}T^2_{\text{DR}}\left[\text{ln}\left(\frac{3}{(N+\frac{1}{2}N_f)}\right)+ \text{ln}(g_d^{-2})+3+2\,\text{ln}(4\pi)-24\,\text{ln}\left( A \right)\right]\,,
\end{split}
\ee
where $A$ is known as the Glaisher-Kinkelin constant given by $\ln A=1/12-\zeta_R'(-1)$ where $\zeta_R$ is the Riemann zeta function, such that $A\simeq 1.28243$.

The third term in the integrand of~\eqref{eq:Gamma1unintegrated} cannot be done analytically. We start to deal with it by making some simplifications
\begin{eqnarray}
	    iii={}&&C m_{\chi}^2\int \frac{p_1^2\text{d}p_1}{2\pi^2}\frac{\partial f_{\text{DR}}^{(0)}(p_1)}{\partial p_1}\,\text{ln}\Big(\frac{m_{\text{D}}^2}{4p_1^2}+1\Big)\\
	={}&&-C m_{\chi}^2\frac{T^2_{\text{DR}}}{6} \int_0^{\infty} \frac{\text{d}x}{2\pi^2}6f_{\text{DR}}^{(0)}(x)\left[2x\,\text{ln}\Big(\frac{(2N+N_f)g_d^2}{24}\frac{1}{x^2}+1\Big)-\frac{2 (2N+N_f)g_d^2x }{(2N+N_f)g_d^2+24x^2} \right]\nonumber\\
	\equiv{}&&-C m_{\chi}^2\frac{T^2_{\text{DR}}}{6}\,\text{F}(g_d)\,.\nonumber
\end{eqnarray}
In the second step we used the change of variable $x=p_1/T_{\text{DR}}$ and integration by part, and set $f_{\text{DR}}^{(0)}(x)=(e^x-1)^{-1}$ in terms of $x$. In the last step we define $\text{F}(g_d)$ as the full term in the integral over $x$.

We are interested in an expansion of $\text{F}(g_d)$ in the coupling $g_d$. However, when naively Taylor-expanding the integrand in powers of $g_d^2$, the individual terms would lead to ill-defined integrals. This means $\text{F}(g_d)$ does not have an analytic expansion in $g_d^2$. Inspecting the integrand we see that for large $x\gg 1$ it is exponentially suppressed due to the DR distribution function $f_{\text{DR}}^{(0)}\to e^{-x}$. However, due to the infrared singularity $f_{\text{DR}}^{(0)}\to 1/x$ for small $x$ (related to Bose enhancement), the integral is sensitive to the integration region $x\lesssim g_d\ll 1$. To make progress, we use the following method to obtain the asymptotic behavior for small $g_d$, inspired by the method of regions~\cite{Beneke:1997zp}. The idea is to split the integral into two parts
\begin{equation}
\text{F}(g_d)  = \text{F}_1^\lambda(g_d) + \text{F}_2^\lambda(g_d)\,.
\end{equation}
First, we perform the integration between $0$ and a certain cutoff $\lambda$, chosen such that $g_d\ll \lambda\ll 1$, allowing us to expand $f_{\text{DR}}^{(0)}$ for small $x$. Second, we perform the integration from this cutoff $\lambda$ to infinity. For this second part we are allowed to expand the integrand in powers of $g_d^2$, since $g_d\ll x$ in the $\lambda \leq x<\infty$ region, while the small-$x$ behaviour is important for $0\leq x< \lambda$. In the end, of course, $\text{F}$ should be independent of the arbitrary parameter $\lambda$.

For the first part F$_1^\lambda$, to study the properties of the asymptotic expansion, we take the expansion of the Bose-Einstein distribution up to the fifth order
into account and we perform the integration from $0$ up to the cutoff $\lambda$,
\begin{eqnarray}\label{eq:F1}
\text{F}_1^\lambda(g_d)={}&&\int_0^{\lambda}\frac{\text{d}x}{2\pi^2}6\, \bigg(\frac{1}{x}-\frac{1}{2}+\frac{x}{12}-\frac{x^3}{720}+\frac{x^5}{30240}+{\cal O}(x^6)\bigg) \times \nonumber \\ &&\bigg[2x\,\text{ln}\Big(\frac{(2N+N_f)g_d^2}{24}\frac{1}{x^2}+1\Big)
-\frac{2 (2N+N_f)g_d^2x }{(2N+N_f)g_d^2+24x^2} \bigg]\,.	
\end{eqnarray}
 We could not analytically compute this integral for an arbitrary constant $\lambda$. That is why we computed it for different values of the cutoff to determine its $\lambda$-dependence behavior. 

\begin{table}[ht]
	\begin{center}
\begin{tabular}{|c| c| c| c|c| c| c| } 
 \hline
  & $g_d$ & $g_d^2$ & $g_d^3$ & $g_d^4$ & $g_d^5$ & $g_d^6$ \\[0.5ex] 
 \hline
$1/x$ & $\frac{\sqrt{2N+N_f}}{2\pi}\sqrt{\frac{3}{2}}$ &  &  & $c^{(1)}_1(\lambda)$ &  & $d^{(1)}_1(\lambda)$  \\ [0.5ex] 
 \hline
 $-1/2$ &  & $-\frac{2N+N_f}{16\pi^2}$ &  & $c^{(1)}_2(\lambda)$ &  & $d^{(1)}_2(\lambda)$  \\[0.5ex] 
 \hline
 $x/12$ &  &  & $\frac{(2N+N_f)^{3/2}}{576 \sqrt{6}\pi}$ & $c^{(1)}_3(\lambda)$ &  & $d^{(1)}_3(\lambda)$ \\[0.5ex] 
 \hline
 $-x^3/729$ &  &  &  & $c^{(1)}_4(\lambda)$ & $\frac{(2N+N_f)^{5/2}}{466560\sqrt{6}\pi}$ & $d^{(1)}_4(\lambda)$  \\[0.5ex] 
 \hline
\end{tabular}
\end{center}
\caption{\label{tab:gexpansion}\small
Contributions to the small-$x$ region $0\leq x <\lambda$ of the function F$_1^\lambda(g_d)$, defined in~\eqref{eq:F1}. {\it Rows:} powers of $x$ in the expansion of $f_{\text{DR}}^{(0)}(x)=(e^x-1)^{-1}$
around zero. {\it Columns:} powers of $g_d$.
}
\end{table}

Keeping all terms appearing in~\eqref{eq:F1}, we summarize the results in Tab.\,\ref{tab:gexpansion}. We find that the result can be written as a Taylor series in $g_d$ instead of $g_d^2$, as one might have naively expected. This behavior is typical for thermal corrections in presence of infrared singularities due to Bose enhancement~\cite{Laine:2016hma}. From Tab.\,\ref{tab:gexpansion} one can see that up to the sixth order in $g_d$ all terms with an odd power are independent of the cutoff. For the even powers, only $g^2_d$ is independent of $\lambda$. We furthermore see that every term in the expansion of the Bose-Einstein distribution (corresponding to the rows in Tab.\,\ref{tab:gexpansion}) contributes at  fourth and sixth order in $g_d$ in a $\lambda$-dependent way\footnote{We also checked that this is also the case for $g_d^{8}$ and higher orders.}. The lower subscript in the coefficients $c^{(1)}_i(\lambda)$ and $d^{(1)}_i(\lambda)$ refers to the term in the expansion of the Bose-Einstein distribution and the upper index refers to the first part of the integral, i.e. the small-$x$ region $0\leq x <\lambda$. By construction, the cutoff dependence of these $g_d^{2k+4}$ terms ($k=0,1,\dots$) has to cancel out with the cutoff dependence of corresponding terms appearing in the second part of the integral. For now, we can compact the result we obtained for the first part as
\begin{eqnarray}
	    \text{F}_1^\lambda(g_d)={}&&\frac{\sqrt{2N+N_f}}{2\pi}\sqrt{\frac{3}{2}}g_d-\frac{2N+N_f}{16\pi^2}g^2_d+\frac{(2N+N_f)^{3/2}}{576 \sqrt{6}\pi}g^3_d+\sum_ic_i^{(1)}(\lambda)g^4_d\nonumber \\&&
	+\frac{(2N+N_f)^{5/2}}{466560\sqrt{6}\pi}g^5_d+\sum_id_i^{(1)}(\lambda)g^6_d+\mathcal{O}(g_d^7) \,.
\end{eqnarray}
For the second part, i.e. the contribution to F$(g_d)$ from the region $\lambda\leq x < \infty$, the expansion is performed in small values of $g_d^2$ (since $g_d\ll\,x$ in this region), giving
\begin{equation}
    \begin{split}
        \text{F}^\lambda_{2}(g_d)={}&\int_{\lambda}^{\infty}\frac{\text{d}x}{2\pi^2}\frac{6}{e^x-1}2x\bigg(\frac{(2N+N_f)^2}{24^2}\frac{g_d^4}{x^4}-\frac{(2N+N_f)^3}{24^3}\frac{g_d^6}{x^6}\bigg)+\mathcal{O}(g_d^8)\\
        ={}&c^{(2)}(\lambda)g_d^4+d^{(2)}(\lambda)g_d^6+\mathcal{O}(g_d^8)\,.
    \end{split}
\end{equation}
Just as in the former part, it was not possible to perform the integral for an arbitrary value of $\lambda$, and we performed the integral for different fixed values to obtain the structure of the result. Since the second part is analytic, Taylor expansion in $g_d^2$ and integration do commute in this case. Since the expansion of the integrand starts at the fourth order in $g_d$, we therefore obtain an expansion of $\text{F}_{2}^\lambda(g_d)$ starting at the same order, and find that they are cutoff dependent. Their cutoff dependence has to cancel those of the corresponding terms in $\text{F}_1^\lambda(g_d)$, and we see that the structure of the result allows for this to be the case (nevertheless one cannot check the cancellation at order $g_d^4$ explicitly because for $\text{F}_1^\lambda(g_d)$ there is a contribution to $g_d^4$ terms at all orders in the expansion in $x$). Still, the result supports the finding that contributions up to order $g_d^3$ in $\text{F}_1^\lambda(g_d)$ have to be cutoff independent, as confirmed by our explicit result. In summary, the splitting into two regions allows us to analytically extract an expansion of the function F$(g_d)$ up to order $g_d^3$ 
\be 
F(g_d)={}\frac{\sqrt{2N+N_f}}{2\pi}\sqrt{\frac{3}{2}}g_d-\frac{2N+N_f}{16\pi^2}g^2_d+\frac{(2N+N_f)^{3/2}}{576 \sqrt{6}\pi}g^3_d+{\cal O}(g^4_d)\nonumber\,.
\ee
After all of the previous discussion, we can conclude that the $t$-channel squared amplitude gets corrections at linear order in $g_d$ and also higher orders that one can safely neglect. We can write this part as
\begin{equation}\label{eq:ii}
    iii={}-C m_{\chi}^2\frac{T^2_{\text{DR}}}{6}\bigg(\frac{\sqrt{2N+N_f}}{2\pi}\sqrt{\frac{3}{2}}g_d+\mathcal{O}(g_d^2) \bigg) \,.
\end{equation}

The fourth term in the integrand of~\eqref{eq:Gamma1unintegrated} is also a bit tricky to deal with, and we use a similar procedure for it as above, 
\begin{eqnarray}\label{eq:G}
 	         iv = &&{} C  m_{\chi}^2 \int \frac{p_1^2\text{d}p_1}{2\pi^2}p_1^2\frac{\partial f_{\text{DR}}^{(0)}(p_1)}{\partial p_1}\left(-\frac{4}{m_{\text{D}}^2+4p_1^2}\right)\nonumber \\
 	= &&{} Cm_{\chi}^2\frac{T_{\text{DR}}^2}{6}\int_0^{\infty}\frac{\text{d}x}{2\pi^2}f_{\text{DR}}^{(0)}(x)\left[\frac{12 x}{\frac{(2N+N_f)}{24}\frac{g_d^2}{x^2}+1}+ \frac{(2N+N_f)g_d^2}{2x\left( \frac{(2N+N_f)}{24}\frac{g_d^2}{x^2}+1\right)^2}\right]\\
 	\equiv &&{} Cm_{\chi}^2\frac{T_{\text{DR}}^2}{6}G(g_d)\,.\nonumber
 \end{eqnarray}
In the second step, we use again the change of variable ($x=p_1/T_{\text{DR}}$), insert the Debye mass~\eqref{eq:debyeM}, and perform an integration by part. In the final step, we define the function $G(g_d)$, which again cannot be computed analytically. To proceed, we use the same logic as before, which leads to  
\begin{equation}
    G(g_d)={}1-\frac{\sqrt{2N+N_f}}{4\pi}\sqrt{\frac{3}{2}}g_d+\frac{9}{16\pi^2}g^2_d-\frac{3}{128\pi}\sqrt{\frac{3}{2}}g_d^3+{\cal O}(g_d^4)\,.
\end{equation}
Inserting this result into~\eqref{eq:G}, we obtain
\begin{equation}\label{eq:iii}
    iv={}Cm_{\chi}^2\frac{T_{\text{DR}}^2}{6}\left(1-\frac{\sqrt{2N+N_f}}{4\pi}\sqrt{\frac{3}{2}}g_d+\mathcal{O}(g^2_d) \right)\,.
\end{equation}

Now we can combine our results in (\ref{eq:i}, \ref{eq:ii}, \ref{eq:iii}) to get the full contribution to the IDM-DM interaction rate from the square of the $t-$channel amplitude,
\begin{eqnarray}\label{eq:gammat}
	     \Gamma_{\rm IDM-DR}^{t}={}&&-Cm_{\chi}^2\frac{T_{\text{DR}}^2}{6}\bigg(\text{ln}(\alpha_d^{-1})+\frac{3\sqrt{2N+N_f}}{4\pi}\sqrt{\frac{3}{2}}g_d+2+\text{ln}\left(\frac{6}{(2N+N_f)}\right)
	     \nonumber \\
	&& +\text{ln}(4\pi) -24\,\text{ln}\,(A)+\mathcal{O}(g_d^2) \bigg)\,.
\end{eqnarray}

The computation of the remaining terms in the squared amplitude in \eqref{eq:totalM} coming from the square of the $s/u$-channel and interference terms (we denote them by $s,u,\dots$) is straightforward
\begin{subequations}
    \begin{align}
        A_0^{s,u,\dots}(p_1)={}& -\frac{2}{3N^2}g_{\text{d}}^4+g_{\text{d}}^4\left[2+\frac{m_{\chi}^2}{p_1^2}+\frac{m_{\chi}^2}{p_1^2}\bigg(\text{ln}\,(m_{\text{D}}^2)-\text{ln}\,(m_{\text{D}}^2+4p_1^2) \bigg) \right]\,,\\
        A_1^{s,u,\dots}(p_1)={}& g_{\text{d}}^4 \left[\frac{2m_{\chi}^2}{p_1^2}+\frac{m_{\chi}^2}{p_1^2}\bigg(\text{ln}\,(m_{\text{D}}^2)-\text{ln}\,(m_{\text{D}}^2+4p_1^2) \bigg) \right]\,,\hspace{1cm} \text{for} \,\,\,p_1\gg m_{\text{D}}\,.
    \end{align}
\end{subequations}
By combining these two results the Debye mass drops out, and the contribution to the IDM-DM interaction rate from the square of the $s/u$-channel and interference terms reads
\be
\Gamma_{\rm IDM-DR}^{s,u,\dots}={}C\frac{4\pi^2}{15}T_{\text{DR}}^4\bigg(\frac{1}{3N^2}-1 \bigg)-C \frac{m_{\chi}^2}{6}T^2_{\text{DR}} \bigg(-1 \bigg)\,.
\ee
The first term is highly suppressed since we are taking the limit where $m_\chi \gg T_{\rm DR}$. The second term gives a further correction to the interaction rate, at the same order as the ${\cal O}(g_d^0)$ terms in the expansion used in~\eqref{eq:gammat}. It arises from the interference of the $t-$channel diagram with the $s-$ and $u-$channel contributions. We can combine it with the previous result from the square of the $t-$channel contribution in~\eqref{eq:gammat} to obtain the full interaction rate of the non-Abelian DM-DR model,
\begin{eqnarray}
\Gamma_{\rm IDM-DR}={}&&-C \frac{m_{\chi}^2}{6}\bigg\{T^2_{\text{DR}}\bigg( \text{ln}(\alpha_d^{-1})+1+\text{ln}\,\left(\frac{6}{(2N+N_f)}\right)+\text{ln}(4\pi) -24\,\text{ln}\,(A)\nonumber \\
{}&&+\frac{3\sqrt{2N+N_f}}{4\pi}\sqrt{\frac{3}{2}}g_d+\mathcal{O}(g_d^2)\bigg)+\mathcal{O}\bigg(\frac{T_{\rm DR}^4}{m_{\chi}^2} \bigg)\bigg\} \,.	
\end{eqnarray}

\bibliographystyle{JHEP}
\bibliography{ref.bib}

\end{document}